\documentclass[aps,prl,twocolumn,superscriptaddress]{revtex4}
\usepackage{subfigure}
\usepackage{graphicx}
\usepackage{rotating}
\usepackage{epstopdf}
\usepackage{amsmath,amsthm}

\begin{document}
\title{Effects of dimethyl sulfoxide on surface water near phospholipid bilayers}
\author{Yuno Lee}
\affiliation{Korea Institute for Advanced Study, Seoul 130-722, Republic of Korea}
\author{Philip A. Pincus}
\affiliation{Physics and Materials Departments, University of California, Santa Barbara, California 93106, USA}
\author{Changbong Hyeon}
\affiliation{Korea Institute for Advanced Study, Seoul 130-722, Republic of Korea}

\date{\today}
\begin{abstract}
Despite much effort to probe the properties of dimethyl sulfoxide (DMSO) solution, effects of DMSO on water, especially near plasma membrane surfaces still remain elusive.  
By performing molecular dynamics (MD) simulations at varying DMSO concentrations ($X_{\text{DMSO}}$), 
we study how DMSO affects structural and dynamical properties of water in the vicinity of phospholipid bilayers. 
As proposed by a number of experiments, our simulations confirm that DMSO induces dehydration from bilayer surfaces and disrupts the H-bond structure of water.    
However, DMSO enhanced water diffusivity at solvent-bilayer interfaces, an intriguing discovery reported by 
a spin-label measurement, is not confirmed in our simulations. 
In order to resolve this discrepancy, we examine the location of the spin-label (Tempo), relative to the solvent-bilayer interface. 
In accord with the evidence in the literature, our simulations, which explicitly model Tempo-PC, find that the Tempo moiety is equilibrated at $\sim 8-10$ \AA\ \emph{below} the bilayer surface. 
Furthermore, the DMSO-enhanced surface water diffusion is confirmed only when water diffusion is analyzed around the Tempo moiety that is immersed below the bilayer surface, which implies that the experimentally detected signal of water using Tempo stems from the interior of bilayers, not from the interface. 
Our analysis finds that the increase of water diffusion below the bilayer surface is coupled to the increase of area per lipid with an increasing $X_{\text{DMSO}}$ $(\lesssim 10\text{ mol\%})$. 
Underscoring the hydrophobic nature of Tempo moiety, our study calls for careful re-evaluation of the use of Tempo in the measurement on lipid bilayer surfaces.
\end{abstract}
\maketitle

\section*{Introduction}
Broadly used in biology as a cosolvent for cryoprotection (mole fraction of DMSO, $X_{\text{DMSO}}\approx 0.1$) and an enhancer of cell fusion and membrane permeability at high concentration ($X_{\text{DMSO}}\gtrsim 0.6$), effects of DMSO on aqueous environment have been a subject of great interest for many decades \cite{Lovelock1959Nature,Mazur1970Science,anchordoguy1991Cryobiology,soper1992JCP}.  
While great progress has been made in understanding how DMSO affects the structural and dynamical properties of bulk water \cite{Luzar93JCP}, the effects of DMSO on water molecules at lipid bilayer surfaces remain elusive.  
Surface forces apparatus (SFA) measurement between two supported DPPC bilayers \cite{Schrader2015PNAS} and the observed decreasing repeat distance of multi-lamellar structures with increasing $X_{\text{DMSO}}$ \cite{Gordeliy1998BJ,kiselev1999dmso} suggest that DMSO dehydrates bilayer surfaces. 
Employing Overhauser dynamic nuclear polarization (ODNP) measurement that used the Tempo (2,2,6,6-tetramethylpiperidin-1-oxyl) moiety as a spin-label to probe the dynamics of water,  
Cheng \cite{cheng2015BJ}  \emph{et al.} reported that water diffusion was enhanced in the vicinity of bilayer surfaces with increasing $X_{\text{DMSO}}$. 
These companion studies \cite{Schrader2015PNAS, cheng2015BJ} argue that both the surface dehydration and enhanced water diffusivity at the bilayer surfaces originate from DMSO-weakened H-bonds of water with lipid head groups.   
Although the properties of surface water at bilayers is known to differ from those of bulk water \cite{bhide2005JCP, Berkowitz2012ACR}, 
the DMSO enhanced-surface water diffusion is both intriguing and counter-intuitive, not easily reconciled with the general notion that DMSO slows down the water dynamics (e.g., diffusion, H-bond lifetime) (see Fig.\ref{bulk_water_dynamics}) \cite{Luzar93JCP,nieto2003JCP}. 
Furthermore, the assumption that a strong interaction of DMSO with lipid head groups displaces hydration water and weakens the H-bond network near bilayer surfaces \cite{yu2000BBA,Gordeliy1998BJ} was surmised based on a few pieces of indirect evidence stitched together from measurements conducted in bulk and force-distance profiles obtained by SFA \cite{Schrader2015PNAS}, not based on a direct measurement probing the dynamics of water and DMSO on surfaces. We believe that this requires further study and confirmation.

In order to unravel the molecular origin underlying the experimental observations, 
we performed atomistic MD simulations of phopholipid-DMSO-H$_2$O systems at various $X_{\text{DMSO}}$ (see Methods for the details of force fields used). 
In the following, we first calculate the density and diffusivity profiles of water and DMSO at POPC bilayers, visualizing the effects of DMSO on water near bilayer surfaces.  
Next, by explicitly modeling Tempo moieties appended to bilayer head groups, we analyze water dynamics around each Tempo moiety. We discuss our simulation results in comparison with Cheng \emph{et al.}'s spin-label measurements \cite{cheng2015BJ} and point out that the hydrophobic nature of Tempo is currently underestimated in the measurement. 
Much complication arises in data interpretation because Tempo appended to PC headgroup is buried below solvent-bilayer interfaces.

\section*{Methods}
{\bf Simulations.} 
MD simulations of palmitoyloleoylphosphatidylcholine (POPC) bilayers were performed with two different cosolvents (DMSO and sucrose). 
The simulation system was constructed with 128 POPC and 4740 to 11664 water molecules, the number of which varied depending on DMSO concentration (Table \ref{table:table_systems}). All the simulations were performed using GROMACS software (ver.4.5.4) \cite{berendsen1995,pronk2013}.
The starting simulation box size was $\sim 6\times 6\times 14$ nm$^3$ with periodic boundary condition, and the system was neutralized with $\sim 50$ Na$^+$ and Cl$^-$ ions corresponding to $\sim$ 150 mM salt concentration. 
The unfavorable inter-atomic contacts in the initial configurations were relieved by the steepest descent energy minimization.  
The system was subjected to position-restrained runs for 1 ns under the NVT ensemble at 300K, followed by 5 ns equilibration run under the NPT (P = 1 bar) ensemble. 
The temperature and pressure was semi-isotropically controlled by Nos{\'e}-Hoover thermostat (with a coupling constant of $\tau_T$ = 0.5 ps) and Parrinello-Rahman barostat (with a coupling constant of $\tau_{P(x-y)} = \tau_{P(z)}$ = 2 ps), respectively.
The cutoff value of 12 \AA\ was used for the both short-range van der Waals and electrostatic interactions. 
For long range electrostatic potential, we used the Particle Mesh Ewald method. 
Each system was simulated for $0.4-1$ $\mu$s and 
the last $0.3-0.9$ $\mu$s of simulation was used for analysis.
\\

\noindent{\bf Force fields for lipid, water, DMSO, sucrose, and Tempo. } 
In reference to the comparative study of PC force field by Piggot et al. \cite{piggot2012}, 
we selected the Berger united atom lipid force field \cite{berger1997} as it can best reproduce the structural and dynamical properties (e.g., area per lipids, volume per lipid, isothermal area compressibility, headgroup-headgroup distance, diffusivity of lipid) of POPC bilayers.  
For water and DMSO, we adopted the simple point charge (SPC) model \cite{hermans1984} and the rigid united-atom model \cite{geerke2004} in the GROMOS 53a6 force field \cite{oostenbrink04JCC}, respectively.  
The topology of sucrose (database identifier: 0ZQQ) generated by B3LYP/6-31G* optimized geometry was obtained from repository of Automatic Topology Builder (ATB) site. 
Next, to model the Tempo moiety, we employed the same Lennard-Jones (LJ) parameters used in modeling POPC. 
Other parameters associated with bond length, angle, and partial charges of the Tempo were adopted from the Ref.\cite{kyrychenko2014} which  conducted density functional theory calculations at the level of UB3LYP/cc-pVDZ. 
\\

\noindent{\bf Density profile of water and cosolvent.}
We counted the number of molecules $N(z,t)$ between $z-\Delta z/2$ and $z+\Delta z/2$ with $\Delta z= 1$ \AA, 
and divided it by a volume $V(t)=x(t)\times y(t)\times \Delta z$ where $x(t)$ and $y(t)$ are the length and width of 
the simulation box. 
The time dependent number density $\rho(z,t)=N(z,t)/V(t)$ was averaged over time, i.e., 
$\rho(z)=\frac{1}{T}\int^T_0\rho(z,t)dt$. 
The number density profiles were plotted for water, DMSO, and sucrose.   
\\

\noindent{\bf Tetrahedral order parameter.} 
The extent of H-bond network formed among water molecules, which is critical to understand the  water structure near bilayer surfaces as well as bulk, is assessed by evaluating the local tetrahedral order parameter \cite{Errington2001Nature,kumar2009PNAS} averaged over the water molecules at position $z$, 
\begin{equation}
\langle Q\rangle(z)=\frac{1}{N(z)}\sum_{k=1}^{N(z)}\left\{1-\frac{3}{8}\sum_{i=1}^3\sum_{j=i+1}^4\left[\cos{\psi_{ikj}}+\frac{1}{3}\right]^2\right\}.
\label{eqn:Q}
\end{equation}
where $N(z)$ is the number of particles present at $z$, $i$ and $j$ denote the nearest neighbors to the water molecule $k$, and $\psi_{ikj}$ denote the angle between the water molecules $i$, $k$, and $j$. 
\\

\noindent{\bf Local diffusivity.}
The local diffusivity is calculated using the finite difference expression \cite{lounnas1994BJ}, which is especially useful for calculating the position-dependent diffusion coefficient in \emph{anisotropic} space.  
\begin{equation}
D=\frac{\langle(\vec{r}(t_2)-\vec{r}(t_0))^2\rangle-\langle(\vec{r}(t_1)-\vec{r}(t_0))^2\rangle}{6(t_2-t_1)}
\label{eqn:D}
\end{equation}
where $\vec{r}(t)$ is the position of a probed atom at time $t$, and $\langle\cdots\rangle$ denotes the average over the ensemble. 
For the time interval of this calculation, we selected $t_2-t_1=1$ ps, 
so that the bulk water diffusion constant in isotropic space calculated from Eq.\ref{eqn:D} quantitatively agrees with the diffusion constant calculated based on the mean square displacement at long time limit, i.e., 
$D\equiv \lim_{t\rightarrow\infty}\langle R^2(t)\rangle/6t$.  
To obtain $z$-dependent diffusion constant ($D(z)$) across bilayer membrane, 
we evaluate Eq.\ref{eqn:D} for the ensemble of molecules in the volume between $z-5$ \AA\ and $z+5$ \AA\ for a given $z$.  

Although subdiffusive behavior of surface water is expected at longer time scale \cite{Berkowitz2012ACR}, 
use of Eq.\ref{eqn:D} should be acceptable for the purpose of comparing our simulation result with experiment, 
as the experiment also estimates  the diffusivity of solvent from short time scale dynamics around spin-labels. 

Some cautionary words are in place. 
Although the bulk water diffusion constant $D_{\text{w}}^{\text{bulk}}= 41\times 10^{-10}$ m$^2$/s obtained from our simulation at $T=300$ K ($=27$ $^oC$) 
is comparable to 
the value $42\times 10^{-10}$ m$^2/$s calculated by Rahman and Stillinger \cite{Rahman1971JCP}, 
it is still $\sim 1.7$ fold greater than the self-diffusion constant of water $D_{\text{w,exp}}^{\text{bulk}}=23\times 10^{-10}$ m$^2$/s measured at 25 $^oC$ \cite{mills1973JPC}. 
It is well known \cite{Mark02JPCA} that all-atom MD simulations using currently available force field generally overestimate (underestimate) the self-diffusion constant (viscosity) of bulk water. 
Thus, when we compare the water diffusion constants calculated from simulations directly with those from experiments, we will take this difference into account by multiplying a correction factor $\phi=D_{\text{w,exp}}^{\text{bulk}}/D_{\text{w}}^{\text{bulk}}\approx 0.56$. 
\\

\noindent{\bf Potential of mean force of the Tempo moiety across lipid bilayers.}
Umbrella sampling technique was used to calculate the free energy of Tempo moiety across the POPC lipid bilayer. 
(i) An initial simulation to generate an initial structure for each window run was conducted for $\sim$ 3 ns.
We pulled the oxygen radical of Tempo moiety appended to Tempo-PC along the z-axis by using a harmonic potential $w_i(z)=(k_o/2)(z-vt)^2$ with $v=0.01$ \AA/ps and $k_o$ = 100 kJ/(mol$\cdot$\AA$^2$). 
(ii) Total 16 window runs, which cover the range of $-30{\text{ \AA}}< z < 0$ \AA, were performed using the initial structures generated from the procedure in (i) and umbrella sampling at each window was conducted for 10 ns.
Umbrella potentials were placed every 2 \AA\ with the strength of $k$ = 0.35 kJ/(mol$\cdot$\AA$^2$). 
The strength of the harmonic umbrella potential $k$ was chosen, such that the positional variance of Tempo moiety $\sigma^2$ from the simulation at 0 mol\% DMSO simulation satisfies $k \approx k_BT/\sigma^2$.
\\

\noindent{\bf Analysis of solvent diffusion around the Tempo moiety.} 
To analyze the diffusion of water around the Tempo moiety, we set the absorbing boundary condition at a distance $R$ from the nitroxide radical oxygen of Tempo-PC, and calculated the escape time (first passage time) of water molecules 
from the interior of a sphere ($r<R$). 
Because the spin-spin interaction is dipolar ($\sim -1/r^6$) in nature, 
the signal from spin-label measurement should reflect a stronger correlation with a water molecule initially closer to the nitroxide oxygen. 
In order to include this effect into our estimate of the average escape time from the Tempo moiety, 
we employed a weighting factor 
\begin{equation} 
  w(r_{\alpha}) = \left\{
 \begin{array}{lr}
 1  & \mbox{for } r<\sigma \\
 \frac{e^{(\sigma/r_{\alpha,i})^6}-1}{e-1} & \mbox{for } r\geq \sigma 
 \end{array}
\right.
\label{eqn:simple_one} 
\end{equation}
which decays from 1 to 0. 
We weighed the escape time $\tau_{\alpha}$ for $\alpha$-th water molecule with  $w(r_{\alpha})$. 
In the expression of $w(r_{\alpha})$, $r_{\alpha,i}$ is the initial position of the $\alpha$-th water molecule;  
$\sigma=3.3$ \AA\ is the position of the 1st solvation shell around the nitroxide oxygen of Tempo for the case of bulk water (Fig.\ref{RDF_Tempo_O_vs_water_O}a), 
and we chose $\sigma=5.8$ \AA\ for the water around Tempo moieties that are buried inside bilayer lipid, so that total number of water around Tempo in both calculations is equal to each other (see Fig.\ref{RDF_Tempo_O_vs_water_O}b).  
We calculated the survival probability of water molecule around the nitroxide radical oxygen using $S(t)=1-\int^t_0p(\tau)d\tau$ where $p(t)=N^{-1}\sum_{\alpha=1}^N\delta(t-t_{\alpha})$ with $t_{\alpha}=w(r_{\alpha,i})\tau_{\alpha}$. 
Finally, the average escape time of water from the spin-label was calculated from $\langle\tau\rangle=\int_0^{\infty}S(t)dt$ (see Fig.\ref{FigS_trap_water}). 

In order to calculate the bulk diffusion constant of water, we conducted independent simulations also by explicitly considering the free radical Tempo in the solution. 
In this case, Tempo can also diffuse, thus the diffusion constant of water was calculated by subtracting the contribution of Tempo as $D^{T,\text{bulk}}_{\text{w}}=D^{\text{bulk}}_{\text{sum}}-D^{\text{bulk}}_{\text{Tempo}}$. 
The condition $R=4.5$ \AA\ was used for calculating the $D^{\text{bulk}}_{\text{sum}}$ \cite{armstrong2009JACS}. 
The DMSO-dependent bulk water diffusion constants, determined from our simulations, compare well with those from experiment using Tempo (see Fig.\ref{Comparison_DTempo_three_methods}). 
\\

\section*{Results}
\noindent{\bf Comparison with the previous MD simulation studies.}
Although the effects of DMSO on lipid bilayer had been studied in a number of MD simulation studies   \cite{smondyrev1999BJ,hughes2012JPCB,Sum2003BJ,notman2006JACS,gurtovenko2007JPCB}, these studies were criticized \cite{yu2000BBA,cheng2015BJ}   because the simulation time was too short or the DMSO-engendered instability of bilayer observed in these simulations were inconsistent with experimental observations.
Our bilayer systems at varying DMSO concentrations remained stable.  
Both the area per lipid (APL) and the bilayer thickness, quantified by the head group (phosphate-phosphate) distance ($d_\text{PP}$), were stably maintained throughout the simulation time ($\sim \mu$sec) (Table S1). 
In the simulation study reporting the DMSO-induced pore formation in a bilayer, Hughes \emph{et al.} \cite{hughes2012JPCB} used the same DMSO force field with ours (GROMOS 53a6), but a different force field (GROMOS 54a7) for the lipid, -- we used the Berger united atom lipid force field \cite{berger1997} -- and conducted the simulations at a temperature $T=350$ K, which was 50 K higher than our study ($T=300$ K). 
For much longer simulation times ($\sim$ 1 $\mu$sec) than others, we do not observe a substantial deposition of DMSO into bilayers or a formation of water pores in the bilayers even at high DMSO concentrations ($X_{\text{DMSO}}\simeq 0.3$), in accord with experiments \cite{cheng2015BJ}.   
\\

\begin{figure}[ht]
\centering
\includegraphics[width=1.0\columnwidth]{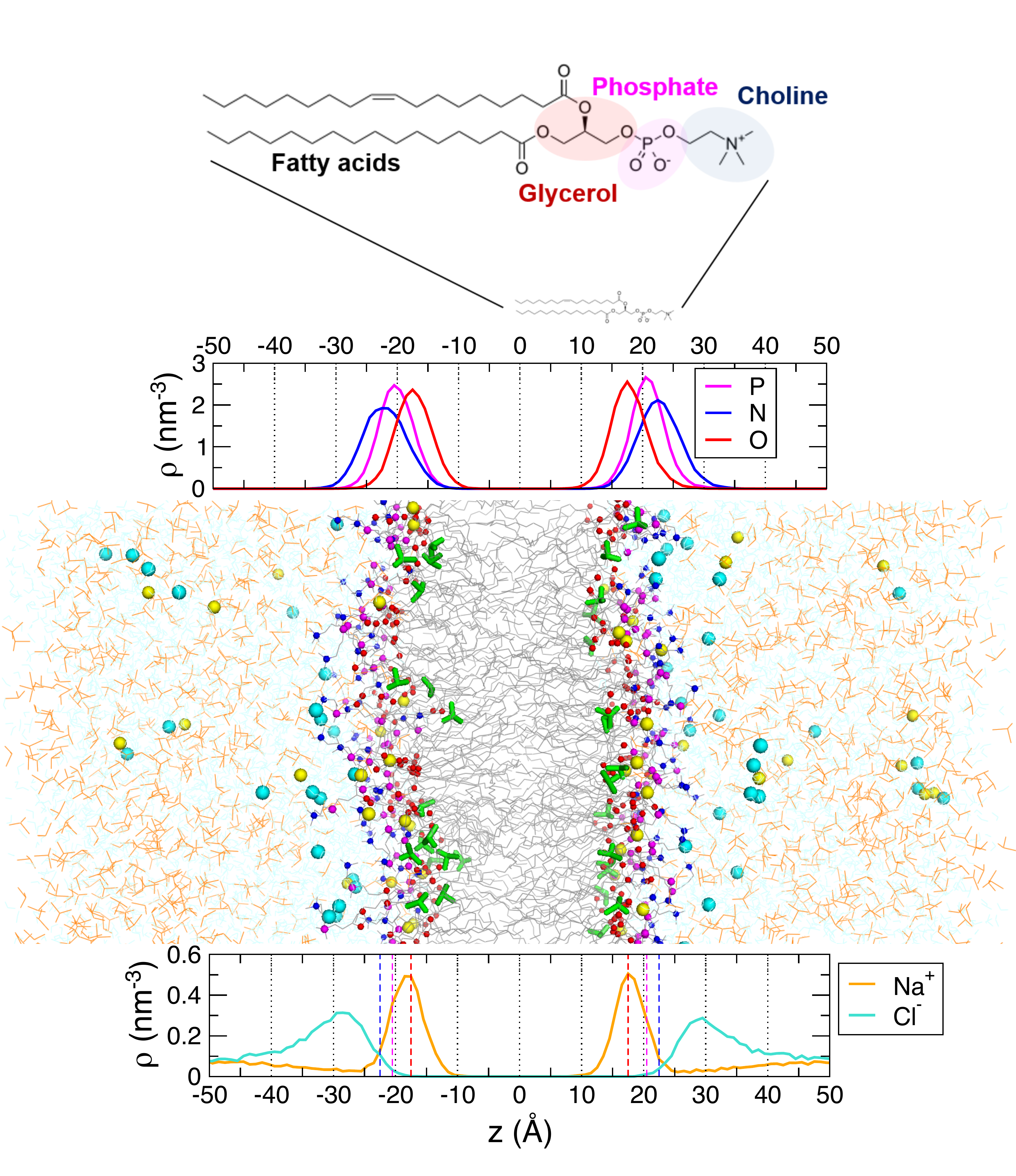}
\caption{A snapshot from MD simulations (at 11.3 mol\% DMSO concentration) performed by solvating POPC bilayers with water, DMSO (orange line), and 150 mM Na$^+$ (yellow spheres), Cl$^-$ (cyan spheres). 
Lipid bilayers are at the center ($z=0$ \AA), and the solvents (water, DMSO, and ions) are at $|z|\gtrsim 20$ \AA.  
DMSO molecules that penetrate inside the headgroups are shown in green sticks. 
On the top, the locations of the phosphate, choline, and glycerol groups in reference to the center of bilayers ($z=0$) are calculated in terms of the number densities of phosphorus (magenta), nitrogen (blue), and oxygen (red) atoms, respectively. 
At the bottom shown are the number densities of Na$^+$ (orange) and Cl$^-$ (cyan) ions across bilayers. 
\label{Fig_system}}
\end{figure}

\begin{figure}[ht]
\centering
	\includegraphics[width=0.95\columnwidth]{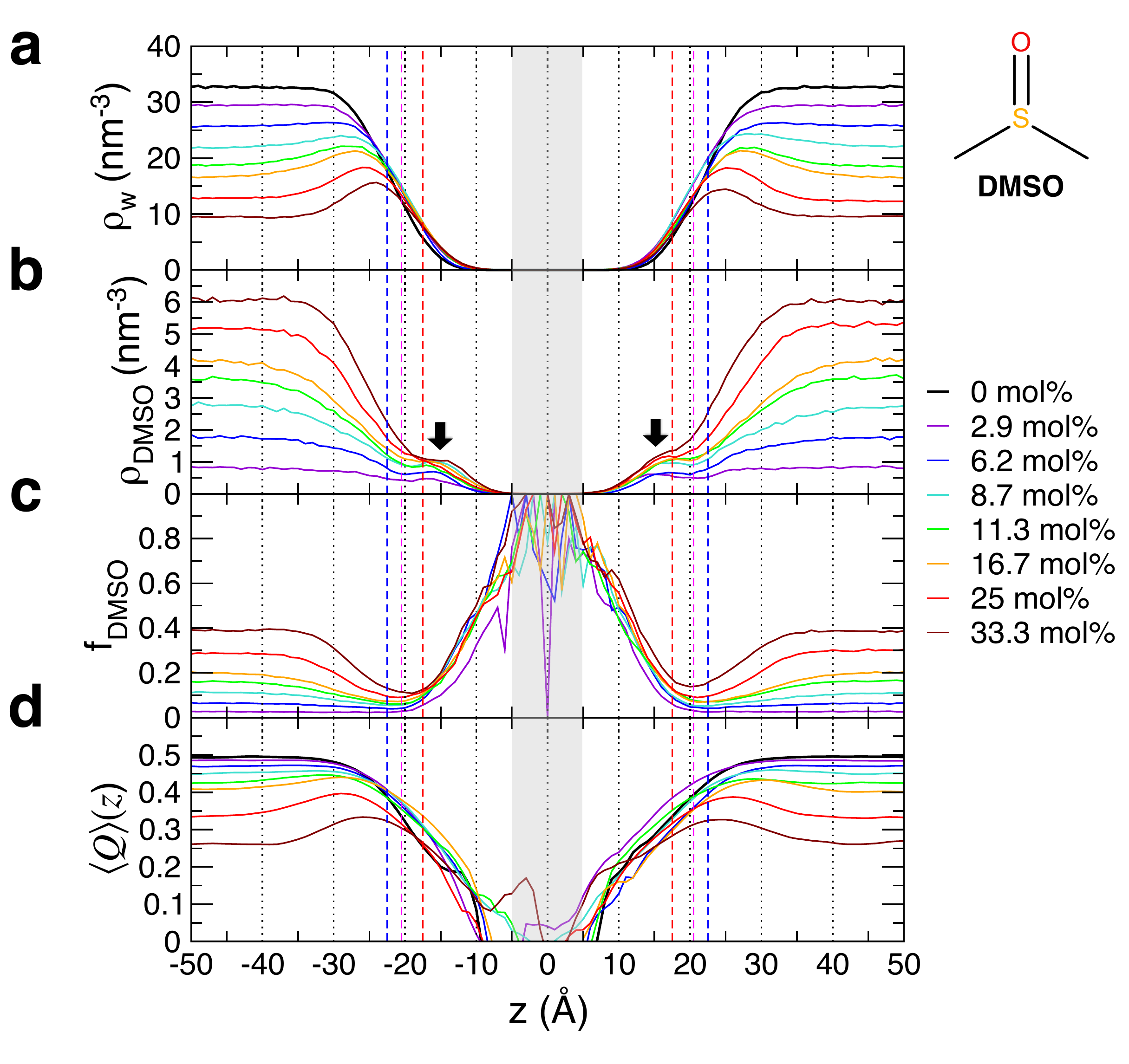}
	\caption{Density profiles of {\bf (a)} water and {\bf (b)} DMSO across POPC bilayers at various DMSO concentrations. 
	{\bf (c)} The proportion of DMSO, $f_{\text{DMSO}}(z)=\rho_{\text{DMSO}}(z)/(\rho_{\text{DMSO}}(z)+\rho_{\text{w}}(z))$. 
	The noisy profile at $|z|\lesssim 5$ \AA, shaded in gray, is due to the lack of statistics in the interior of bilayers, thus should be ignored.   
	{\bf (d)} Tetrahedral order parameter. 
	The dashed lines in red, magenta, and blue are the most probable positions of glycerol, phosphate, and choline groups calculated in Fig.\ref{Fig_system}.    
		\label{Fig_DMSO}}
\end{figure}

\begin{figure}[ht]
\centering
\includegraphics[width=0.95\columnwidth]{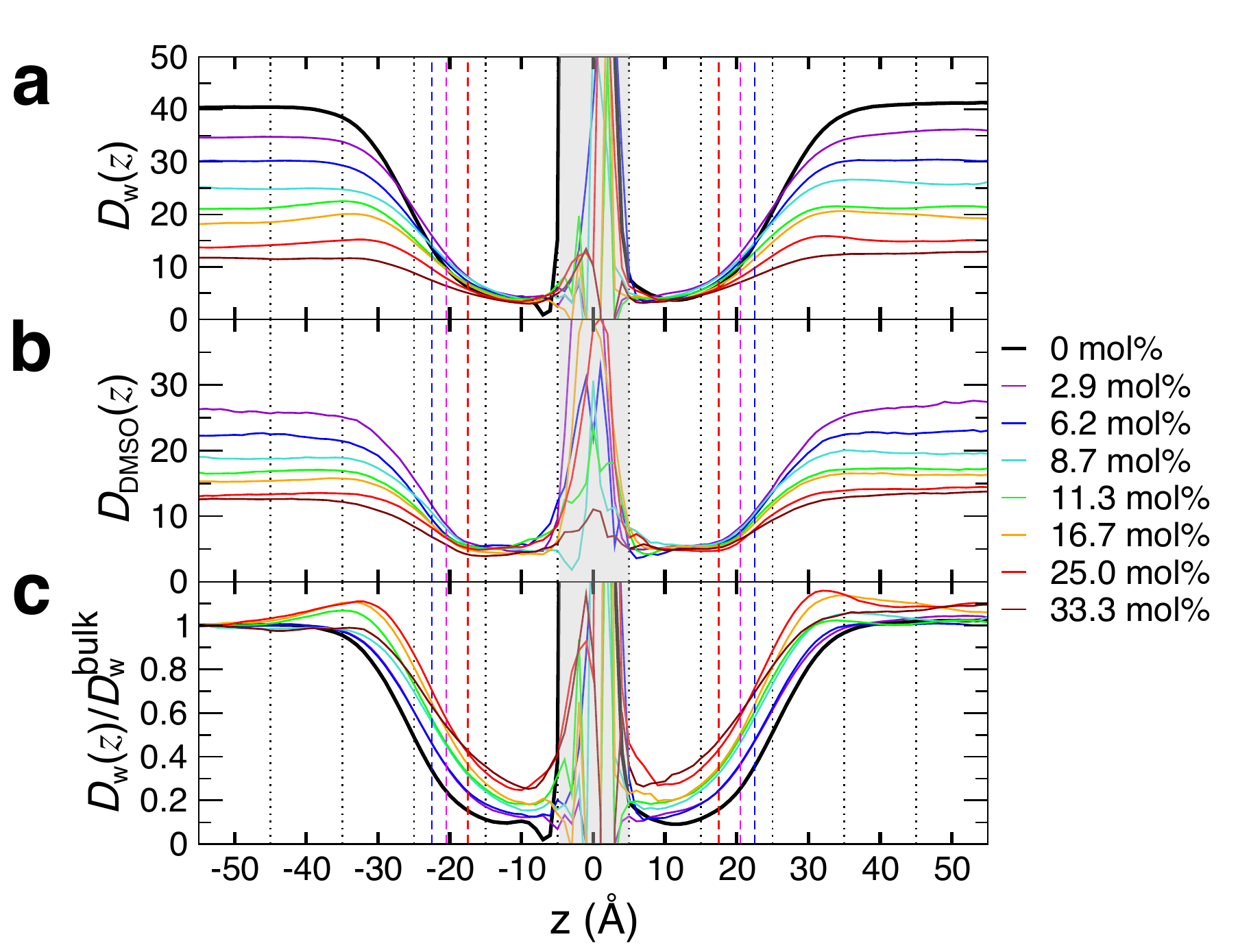}
\caption{Local diffusivities ($\times 10^{-10}$ m$^2$/s) of {\bf (a)} water and {\bf (b)} DMSO across the bilayers. 
{\bf (c)}  Water diffusivity normalized by the value in the bulk, $D_{\text{w}}^{\text{bulk}}\equiv D_{\text{w}}(z=-55{\text{ \AA}})$. 
The dashed lines in red, magenta, and blue are the most probable positions of glycerol, phosphate, and choline groups, respectively.  
The noisy profiles at the bilayer core shaded in gray ($|z|<5$ \AA), due to the paucity of water or DMSO molecules, should be ignored.   
\label{Diffusion}}
\end{figure}

\begin{figure*}[ht]
\centering
\includegraphics[width=1.4\columnwidth]{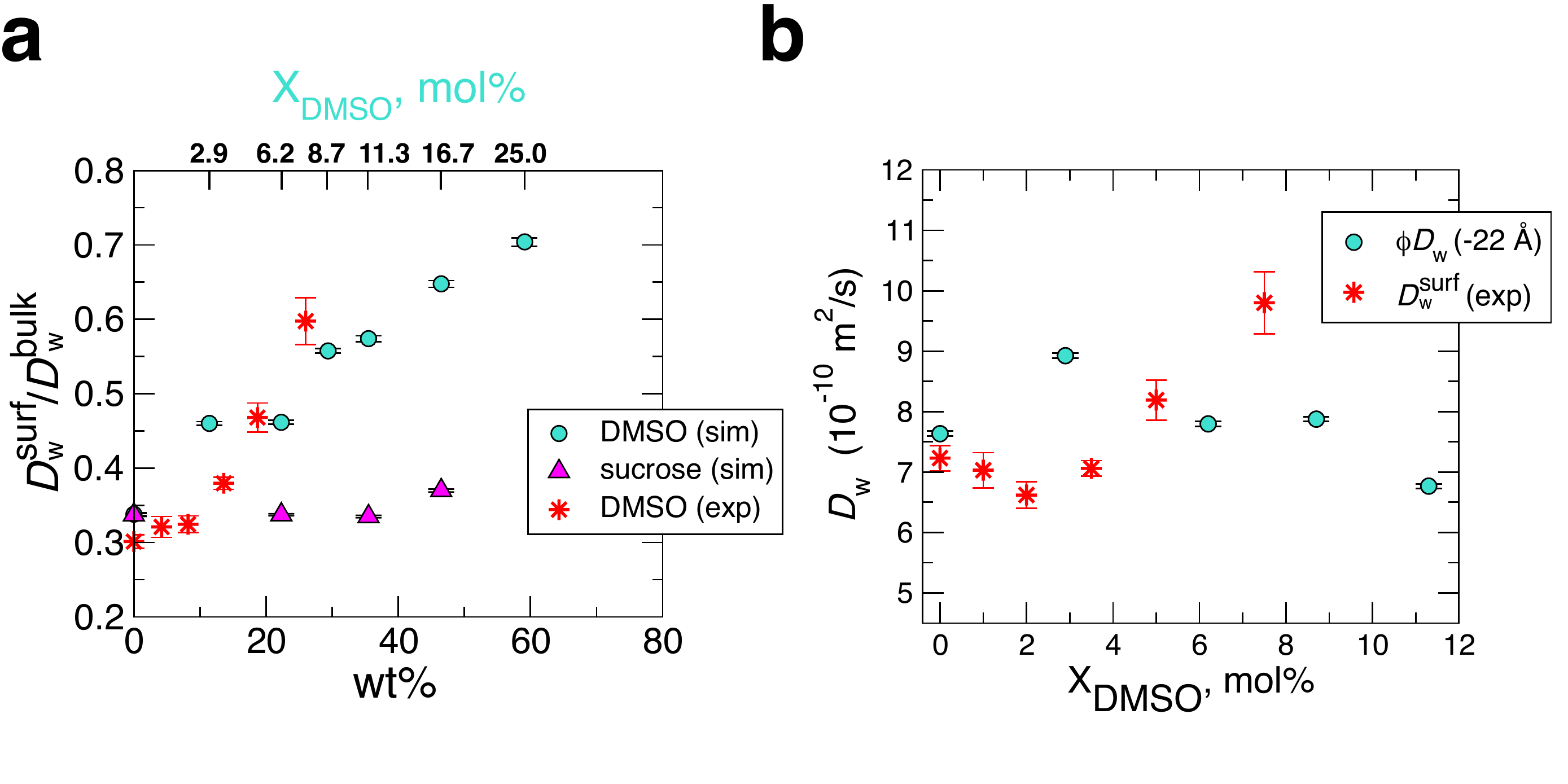}
\caption{Surface water diffusion. 
(a) The diffusion constant, probed at $z=-22$ \AA, i.e., $D_{\text{w}}^{\text{surf}}\equiv D_{\text{w}}(-22{\text{ \AA}})$ 
and normalized with $D_{\text{w}}^{\text{bulk}}$ at each cosolvent concentration (wt\%, the corresponding DMSO mole fraction is annotated at the top), is plotted. 
The data (red asterisks) from the recent ODNP measurement \cite{cheng2015BJ} are overlaid for comparison. 
(b) Diffusion constants of surface water probed at $z=-22$ \AA\ are compared with those from spin-label measurement in Ref.\cite{cheng2015BJ}. 
		To compare our simulation results directly with experiments (Table-S2 (LUV) of Ref.\cite{cheng2015BJ}), the correction factor $\phi=0.56$ (see Methods) was multiplied to $D_{\text{w}}(-22{\text{ \AA}})$. 
		\label{DMSO_diffusivity}}
\end{figure*}

\begin{figure*}[ht]
\centering
	\includegraphics[width=1.8\columnwidth]{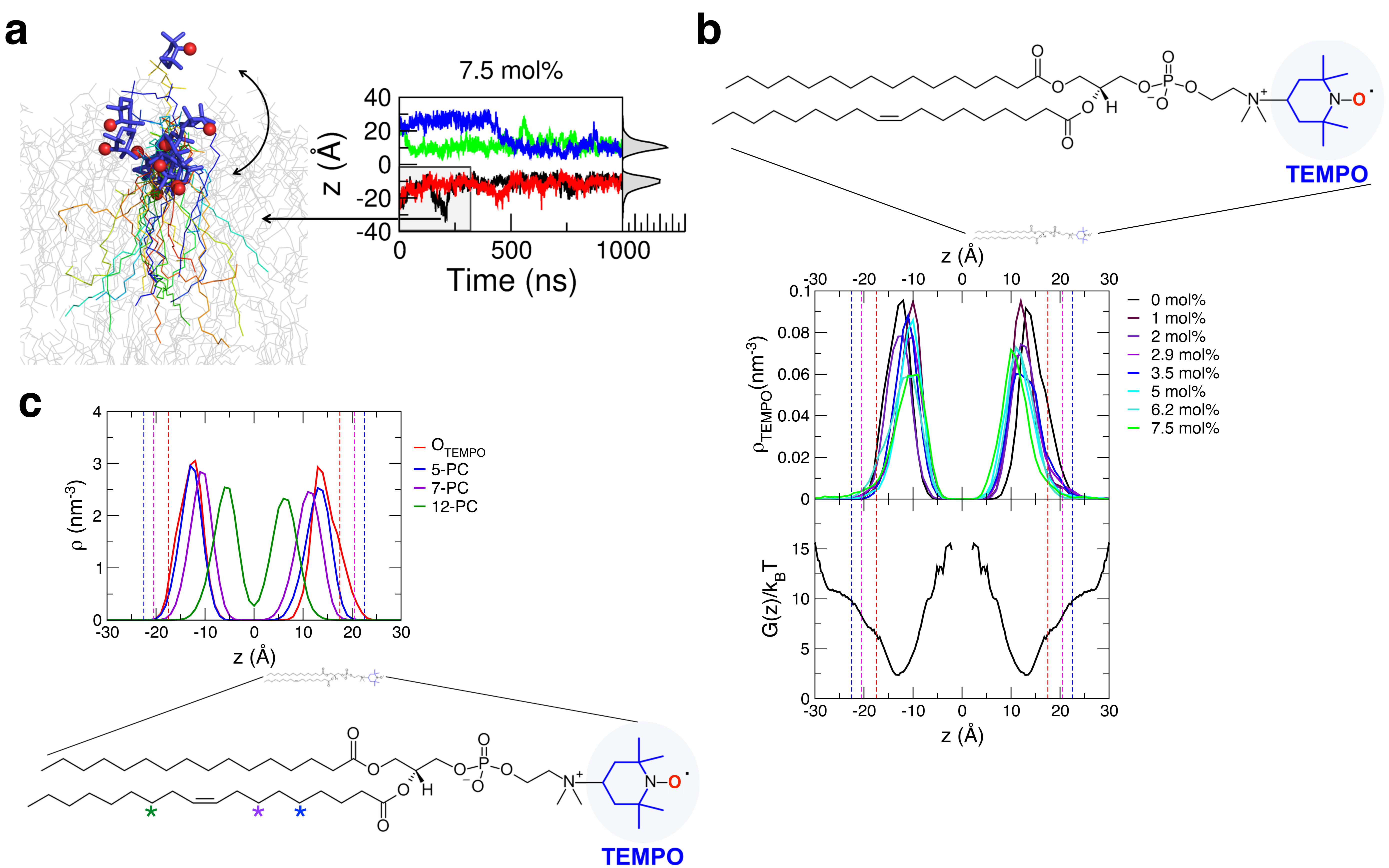}
	\caption{
		\footnotesize
    Dynamics of Tempo moiety.  
    (a) The multiple snapshots of the Tempo-PC from 0 to 300 ns using the black trajectory on the right are overlaid to depict the positions of the Tempo moiety. The Tempo is highlighted with stick representation in blue. The nitroxide radical oxygen is depicted in red sphere.     
    (b) (upper panel) Distribution of nitroxide radical oxygen along the $z$-axis after the position reaches a steady state value. 
    The corresponding time trajectories at varying DMSO concentrations are shown in Fig.\ref{Tempo_pc_z_trace_with_histo}. 
    The equilibrium positions of headgroup (P, N, O) are marked with dashed lines, highlighting that the equilibrated position of Tempo moieties are buried inside from the bilayer surface. 
    The structure of Tempo-PC are shown with highlight on Tempo moiety consisting of 9 hydrocarbons. 
    (lower panel) The PMF of the nitroxide radical oxygen along the $z$-axis, $G(z)/k_BT$, obtained from the umbrella sampling analysis. 
    (c) The distribution of $n$-th hydrocarbon in lipid tail when bilayer lipids are equilibrated. 
		The equilibrated positions of Tempo in the 5-PC, 7-PC, 12-PC, and Tempo-PC are displayed. 
		Each position of $n$-th hydrocarbon is highlighted using the star symbol in the  structure of Tempo-PC.
    	\label{Fig_Tempo}}
\end{figure*}

\noindent{\bf Density profiles of water and DMSO across lipid bilayer. }
Illustrated in a snapshot from simulations (Fig.\ref{Fig_system}), the MD simulations were performed by solvating the POPC bilayer with water, DMSO, and 150 mM NaCl salt. 
The density profiles are presented along an axis normal to bilayer surface ($z$-axis) at a spatial resolution of $\Delta z=1$ \AA\ by averaging the statistics on the $xy$ plane ($\sim 60\times 60$ \AA$^2$) over simulation times (see Table S1); 
$z=0$ is selected as the center of bilayer. 
With the thickness of the POPC bilayer being $\sim$ 40 \AA\ (Table S1), 
the atoms of lipid headgroup are found at $|z|\approx 20$ \AA. 
The distributions of choline, phosphate, and carbonyl groups along the $z$-axis are shown in Fig.\ref{Fig_system} in terms of the number densities of nitrogen (N), phosphorus (P), and oxygen (O) atoms, whose averages are formed at $\langle z_{\text{N}}\rangle \approx 22$ \AA, $\langle z_{\text{P}}\rangle \approx 20$ \AA, and $\langle z_{\text{O}}\rangle \approx 17$ \AA\ (depicted with dashed lines in blue, red, magenta in Figs.\ref{Fig_system}, \ref{Fig_DMSO}, \ref{Diffusion}, \ref{Fig_Tempo}, and  \ref{sucrose}),  respectively. 
The number densities of salt ions (Na$^+$, Cl$^-$) are calculated to show the ion-distribution on the lipid bilayer whose headgroups are made of zwitterions (phosphate and choline groups) (see the density profiles at the bottom panel of Fig.\ref{Fig_system}). 
Na$^+$ ions are condensed on the solvent-bilayer interface with its number density maximized at $|z|\approx 17$ \AA, the position corresponding to the carbonyl oxygen of glycerol group, whereas the distribution of Cl$^-$ ions is maximized at $|z|\approx 30$ \AA.

In order to investigate the structure and dynamics of DMSO solution near the bilayer surfaces, we varied the mole fraction of DMSO from $X_{\text{DMSO}}=$ 0 mol\% to 33.3 mol\% (Figs. \ref{Fig_DMSO}, \ref{Diffusion}). 
In the absence of DMSO ($X_{\text{DMSO}}=0$), 
the bulk water density is $\rho_{\rm w}(z) =\rho_{\text{w}}^{\text{bulk}}\approx 33.3$ nm$^{-3}$ at $|z| \gtrsim 30$ \AA\ (Fig.\ref{Fig_DMSO}a, black line), 
which corresponds to the typical value of 1 g/cm$^3$. 
The water density begins to monotonically decrease $\sim$ 10 \AA\ away from the bilayer surfaces and reduces to $0.5\times \rho_{\text{w}}^{\text{bulk}}$ at the interface ($|z|\approx 22$ \AA). 
The density profile of water $\rho_{\text{w}}(z)$ at $X_{\text{DMSO}}=0$ mol\% is described quantitatively using the interfacial density profile derived from Cahn-Hillard equation \cite{Cahn1958JCP,bu2014JPCC}: 
\begin{align}
\rho_{\text{w}}(z)=\frac{\rho_{\text{w}}^{\text{bulk}}}{2}\left[\tanh{\left(\frac{z-z_{int}}{\sqrt{2}\xi}\right)}+1\right]
\label{eqn:Cahn-Hillard}
\end{align}
with $\rho^{\text{bulk}}_{\text{w}}=32.7$ nm$^{-3}$, $z_{int}=21.8$ \AA, and $\xi=3.75$ \AA, where $z_{int}$ and $2 \xi$ are the position and width of interface, respectively.    
Overall, a decrease of water number density is clearly seen on bilayer surfaces with increasing $X_{\text{DMSO}}$ (Fig.\ref{Fig_DMSO}a); thus confirming the experimentally detected \emph{DMSO induced dehydration} of bilayer surfaces.

Notably, at high DMSO concentrations ($X_{\rm DMSO} \geq$ 8.7 mol\%), 
$\rho_{\rm w}(z)$ displays \emph{non-monotonic} variation from the bulk to interface. 
$\rho_{\text{w}}(z)$ is maximized at $\sim$ $5 - 10$ \AA\ away from the headgroup position ($|z|\approx 22$ \AA), satisfying 
$\rho_{\text{w}}(|z|\approx 25-30 {\text{ \AA}})>\rho_{\text{w}}^{\text{bulk}}$. 
This characteristic ``water rich layer'' near the bilayer surface is specific to the DMSO solution, 
not observed in sucrose solutions (compare $\rho_\text{w}(z)$'s in DMSO with those in sucrose solution in Fig.\ref{sucrose}a).

Unlike $\rho_{\text{w}}(z)$, a ``DMSO rich layer'' above the interface is absent in $\rho_{\rm DMSO}(z)$. 
Instead, a small population of DMSO molecules, depicted in Fig.\ref{Fig_system} with green sticks beneath the headgroup contribute to the small density humps at $|z|\approx 15$ \AA\ (Fig.\ref{Fig_DMSO}b, black arrows). 
At high $X_{\rm DMSO}$, the extent of DMSO depleted from the surfaces (Fig.\ref{Fig_DMSO}b) is greater than that of water, 
which is better demonstrated in the plot calculating the fraction of DMSO, $f_{\text{DMSO}}(z) = \rho_{\text{DMSO}}(z)/(\rho_{\text{DMSO}}(z) + \rho_{\text{w}}(z))$ (Fig.\ref{Fig_DMSO}c).

In order to assess changes in water structure, we calculated the tetrahedral order parameter \cite{Errington2001Nature,kumar2009PNAS} $\langle Q\rangle(z)$ (Eq.\ref{eqn:Q}), which points to an increasing degradation of H-bond network at higher $X_{\text{DMSO}}$ (Fig.\ref{Fig_DMSO}d). 
For a given weight \%, the extent of H-bond network degradation by DMSO is greater than by sucrose (compare Fig.\ref{Fig_DMSO}d and Fig.\ref{sucrose}c. Also, see Fig.\ref{xi_Tetrahedral}b), 
suggesting that compared with sucrose DMSO is a more efficient cryoprotectant.      
\\

\begin{figure}[h!]
	\includegraphics[width=1.0\columnwidth]{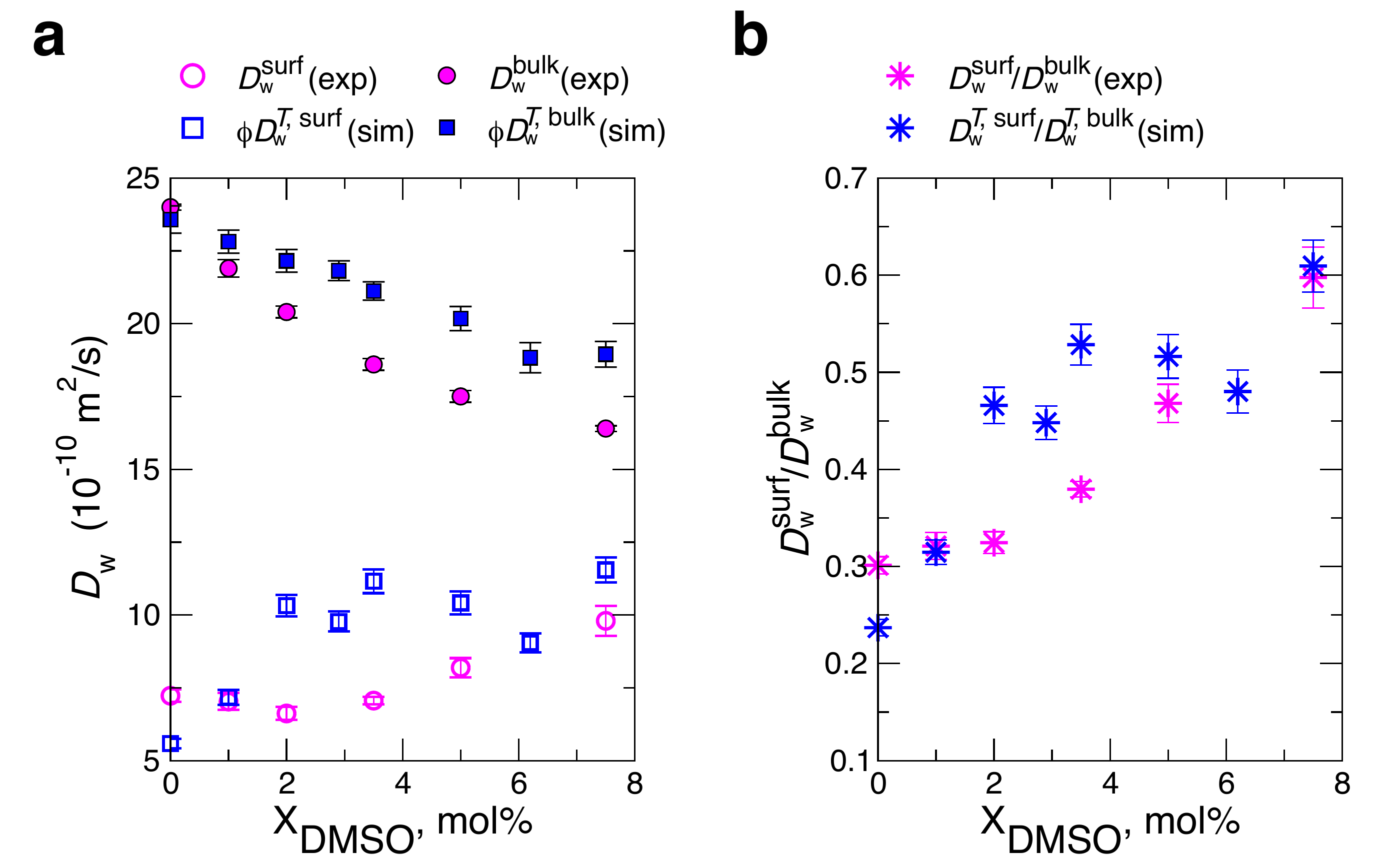}
	\caption{Diffusion coefficients for water molecules {\bf (a)} $D^{\text{surf}}_{\text{w}}$, $D_{\text{w}}^{\text{bulk}}$ from experiment and $D^{T,\text{surf}}_{\text{w}}$, $D_{\text{w}}^{T,\text{bulk}}$ from simulation. 
	{\bf (b)} The ratio $D^{\text{surf}}_{\text{w}}/D_{\text{w}}^{\text{bulk}}$ and $D^{T,\text{surf}}_{\text{w}}/D_{\text{w}}^{T,\text{bulk}}$ at various $X_{\text{DMSO}}$. 
		All the data for simulations were obtained by analyzing water lifetime around Tempo moieties using Eq.\ref{eqn:Einstein}.  
		The experimental values of surface and bulk water diffusion constant are from Table-S2 (LUV) and Table-S3, respectively, of Ref.\cite{cheng2015BJ}. 
			\label{Dsurf_vs_Dbulk}}
\end{figure}

\noindent{\bf Local diffusivity profiles of water and DMSO.}
Local diffusivities of water and DMSO (Fig.\ref{Diffusion}) were calculated using Eq.\ref{eqn:D} \cite{lounnas1994BJ}.  
In the range of $|z|>25$ \AA, which spans the bilayer surface to the bulk, both water and DMSO slow down with increasing $X_{\text{DMSO}}$ (Fig.\ref{Diffusion}a,b). 
While the enhancement of surface water diffusion relative to the bulk water is seen at $|z|\approx 30$ \AA\ (see Fig.\ref{Diffusion}c), $|z|\approx 30$ \AA\ is probably not the position where the spin-label experiment in Ref.\cite{cheng2015BJ} has probed the surface water diffusion. 
The surface water diffusivity is suppressed at  $|z|\approx 22$ \AA, where the experiment tried to probe it using Tempo-PC. 

$D_{\text{w}}^{\text{surf}}/D_{\text{w}}^{\text{bulk}}$ values with varying $X_{\text{DMSO}}$ measured from the ODNP experiment using Tempo spin-labels \cite{cheng2015BJ}, overlaid on the graph (red asterisks in Fig.\ref{DMSO_diffusivity}a), are in good agreement with our simulation result. 
Nevertheless, we argue below that, in contradiction to the original intention, the Tempo-bilayer interaction prevents the Tempo spin-labels from probing the solvent-bilayer interface, $|z|\approx 22$ \AA; thus the agreement of $D_{\text{w}}^{\text{surf}}/D_{\text{w}}^{\text{bulk}}$ between the simulation and experiment in Fig.\ref{DMSO_diffusivity}a is coincidental.  
In the next section, we give an in-depth discussion concerning this issue. 
\\

\section*{Discussion}
\noindent{\bf Analysis of water dynamics around Tempo and its comparison with the spin-label measurement.} 
To experimentally probe the surface water dynamics, 
Cheng \emph{et al.} \cite{cheng2015BJ}  tethered spin-labels (Tempo) to choline groups and conducted measurements 
assuming that the spin-labels remain above or at least near the equilibrium position of choline group ($|z|\approx 22$ \AA). 
They obtained the value of $D^{\text{surf}}_{\text{w}}$ and $D^{\text{surf}}_{\text{w}}/D^{\text{bulk}}_{\text{w}}$  
with varying $X_{\text{DMSO}}$. 
As highlighted in Fig.\ref{DMSO_diffusivity}a, our simulation result of $D^{\text{surf}}_{\text{w}}/D^{\text{bulk}}_{\text{w}}$ is in good agreement with their observations. However, as opposed to the decreasing trend observed in the local diffusivity plot $D_{\text{w}}(z)$ at $z=-22$ \AA\ (overall decrease of $D_{\text{w}}(z=\pm 22\text{ \AA})$ with increasing $X_{\text{DMSO}}$ in Fig.\ref{Diffusion}a), 
Cheng \emph{et al.} reported an overall increase of $D^{\text{surf}}_{\text{w}}$ with $X_{\text{DMSO}}$ ($\ast$ symbols in Fig.\ref{DMSO_diffusivity}b).

In order to investigate the origin of this discrepancy, 
we explicitly modeled Tempo-PCs and placed them at 4 different locations, two Tempo-PCs in the upper leaflet and the remaining two in the lower leaflet of the bilayer. 
The Tempo-PCs are placed $\sim 3$ nm apart from each other to minimize possible interaction between them. 
Since the lateral diffusion constant of lipid molecule on the bilayer surfaces is $\sim 10$ nm$^2/\mu$sec, it is unlikely that two Tempo-PCs actively interact with each other in our simulation time of 1 $\mu$sec. 
Compared with the recent molecular dynamics study of Tempo-PC \cite{kyrychenko2014}, the surface density of Tempo-PC in our study is much lower, precluding an inter-Tempo-PC interaction.

The position of Tempo displays large fluctuations over time, but the major depth distribution of the spin-label was established in the interior of the bilayer ($10<|z|<15$ \AA), instead of at the bilayer surfaces ($|z|\gtrsim 20$ \AA) under all conditions of $X_{\text{DMSO}}=0-7.5$ mol\% (Fig. \ref{Fig_Tempo} and see all the 32 time trajectories, each of which was run for 1 $\mu$sec in Fig.\ref{Tempo_pc_z_trace_with_histo}). 
Furthermore, our simulation time of the bilayer system including Tempo-PC at each DMSO concentration (1 $\mu$sec) is at least five times longer than the previous study \cite{kyrychenko2014}, and the practically irreversible burials of Tempo moieties below bilayer surfaces after the relaxation processes appear to be robust. 
Furthermore, the potential of mean force of Tempo moiety across the POPC bilayer at $X_{\text{DMSO}}=0$ (see Fig.\ref{Fig_Tempo}b, the lower panel), calculated with umbrella sampling technique (Methods), points to the identical location as the most stable location of Tempo and suggests that there is a free energy bias of $\sim 7-8$ $k_BT$ towards the bilayer interior ($|z|\approx 13$ \AA) from the interface ($|z|\approx 22$ \AA), thus giving credence to our simulation results in the upper panel of Fig.\ref{Fig_Tempo}b and Fig.\ref{Tempo_pc_z_trace_with_histo}.  
The hydrophobic nature of Tempo moiety, which is composed of as many as 9 hydrocarbons (see Fig.\ref{Fig_Tempo}b), 
is currently underestimated in the spin-label measurement, but this could be the driving force for this observation. 
We therefore assert that the Tempo moieties probe the interior not the surface of PC bilayer.

Next, in order to make direct comparison of our simulation results with the spin-label experiment, 
we calculated water diffusion around the Tempo using 
\begin{align}
D^T_{\text{w}}=\frac{R^2}{6\langle \tau\rangle}
\label{eqn:Einstein}
\end{align}
where $R=10$ \AA\ for surface \cite{cheng2015BJ} and $R=4.5$ \AA\ for bulk water \cite{armstrong2009JACS}, and the superscript $T$ indicates that the diffusion constant is calculated around Tempo. 
This is the same formula adopted by Cheng \emph{et al.}, and we use the same parameters ($R$) that they use to estimate the diffusion constant of water from ODNP measurement that provide an average spin-spin decorrelation time $\langle\tau\rangle$ \cite{cheng2015BJ} (see Methods for the details of calculating $\langle \tau\rangle$ from simulations). 
We calculated the diffusion constant of water around Tempo moieties after the position of Tempo moiety is equilibrated inside bilayers (Fig.\ref{Tempo_pc_z_trace_with_histo}).  
To make direct comparison of water diffusivity we multiplied the correction factor $\phi\approx 0.56$ (see Methods) to $D_{\text{w}}^{T,\text{surf}}$ and $D_{\text{w}}^{T,\text{bulk}}$ from our simulation.  
The ratio of the two values ($D^{T,\text{surf}}_{\text{w}}/D^{T,\text{bulk}}_{\text{w}}$) are in reasonable agreement with those from experiments over the range of $X_{\text{DMSO}}\leq 8$ \% (Fig.\ref{Dsurf_vs_Dbulk}a), now reproducing the same trends for both surface and bulk water with increasing $X_{\text{DMSO}}$. 
The semi-quantitative agreement of the calculated water diffusion constant around Tempo moieties with Cheng \emph{et al.}'s measurement lends support to our simulation results and the finding that Tempo moieties are equilibrated below the bilayer surfaces. 
\\

\noindent{\bf Equilibrium position of Tempo moiety in PC bilayers.}  
While not explicitly pointed out by the authors, 
evidence of the burial of Tempo moiety inside POPC bilayer surfaces is, in fact, present in the experimental data by Subczynski \emph{et al.} \cite{Widomska2010MMB}. 
They presented oxygen transport parameters, which contains information on the depth of Tempo in terms of the accessibility of nitroxide moiety to the oxygen, across POPC bilayers measured at 25 $^{\circ}C$ (Fig.9b in \cite{Widomska2010MMB}). 
The data indicate that the oxygen transport parameter measured by Tempo-PC is comparable to the parameter measured by 5-PC (POPC lipid with nitroxide label at the position of the 5-th hydrocarbon in lipid tail), which is identified in our simulation at $|z|\approx 13\pm 5$ \AA, approximately the same position where a buried Tempo moiety is equilibrated (Fig.\ref{Fig_Tempo}c).     
The hydrophobicity profile across POPC bilayer, quantified by the $z$-component of hyperfine coupling tensor of nitroxide moiety (Fig.8b in \cite{Widomska2010MMB}) may indicate that Tempo in Tempo-PC is situated in a less hydrophobic environment than the interior of bilayer; however, it should be noted that the measurement in Ref. \cite{Widomska2010MMB} was conducted at a cryo condition ($T=-165$ $^\circ$C). 
The burial of Tempo appended to the PC headgroup has also been reported in the MD simulation studies by Kyrychenko \emph{et al.} \cite{kyrychenko2013JPCB,kyrychenko2014}, who pointed out ``a much broader and heterogeneous distribution for a head-group-attached Tempo spin-label of Tempo-PC lipids,'' alerting ``the possible sources of error in depth-dependent fluorescence quenching studies.''  
Kyrychenko and Ladokhin also experimentally showed that Stern-Volmer constant ($\tau_0/\tau_Q$), corresponding to the inverse of fluorescence quenching time, ($\tau_Q$), of NBD-PE increases in the order of 
12-Doxyl-PC$<$Tempo-PC$<$7-Doxyl-PC$<$5-Doxyl-PC, suggesting that Tempo-moiety in Tempo-PC is more deeply buried than 5- or 7-Doxyl-PC \cite{Ladokhin14AB}. 
The hydrophobic nature of the Tempo moiety and inherent disorder caused by thermal motion \cite{Ladokhin14AB}, the latter of which is especially relevant for Tempo-PC, should be taken seriously.  
\\

\noindent{\bf Thermodynamic and kinetic effects of DMSO on surface water.}
Now that we have reproduced the trend of surface water diffusivity measured by the spin-label experiment semi-quantitatively, we are in a good position to examine the mechanistic proposals made in the two reports \cite{Schrader2015PNAS,cheng2015BJ}. 
The experimental studies using SFA and ODNP measurements \cite{Schrader2015PNAS,cheng2015BJ} observed (i) a decrease of membrane repulsion \cite{yu1998BiophysChem} and (ii) enhanced local diffusivity of surface water in the presence of DMSO ($X_{\text{DMSO}}<10$ mol\%), respectively. 
The authors surmised 
that the competition between DMSO and surface hydration water on the interaction with lipid head-groups led to weakening of the strength of the cohesive water network hydrating the membrane head-groups, and thus enhancing surface water diffusivity and dehydration of the bilayer surfaces.
Our computational study on lipid-DMSO-H$_2$O system not only reproduces these observations, but it also provides more accurate understanding of what is actually happening on the bilayer surfaces at the molecular level by making accessible the profiles of density, local diffusivity, and tetrahedral order parameter for water structure:  
\\

\noindent{\it Water structure}: 
Density profiles ($\rho_{\text{w}}(z)$) of water across lipid bilayers visualize the dehydration due to DMSO (Fig.\ref{Fig_DMSO}a). 
The tetrahedral order parameter ($\langle Q\rangle(z)$) indicates that the hydrogen bond network 
begins to be disrupted $\sim$ 10 \AA\ away from the interface ($|z|=z_{int}$) even in the absence of DMSO, and 
the degradation of H-bond network is promoted at cosolvent concentrations (Figs.\ref{Fig_DMSO}d, \ref{sucrose}c, \ref{xi_Tetrahedral}b). 
An interesting finding from our study is that
the extent of change in the surface water density ($\rho_{\text{w}}(z\approx z_{int};X_{\text{DMSO}})$) is smaller than the change of the bulk water density ($\rho_{\text{w}}^{\text{bulk}}(X_{\text{DMSO}})$) (Fig.\ref{Fig_DMSO}a). This leads to a ``water rich layer'' which is specific to DMSO solutions at high $X_{\text{DMSO}}$ ($\gtrsim 8.7$ mol\%) (Fig.\ref{Fig_DMSO}a). 
\\

\begin{figure}[ht]
	\includegraphics[width=1.0\columnwidth]{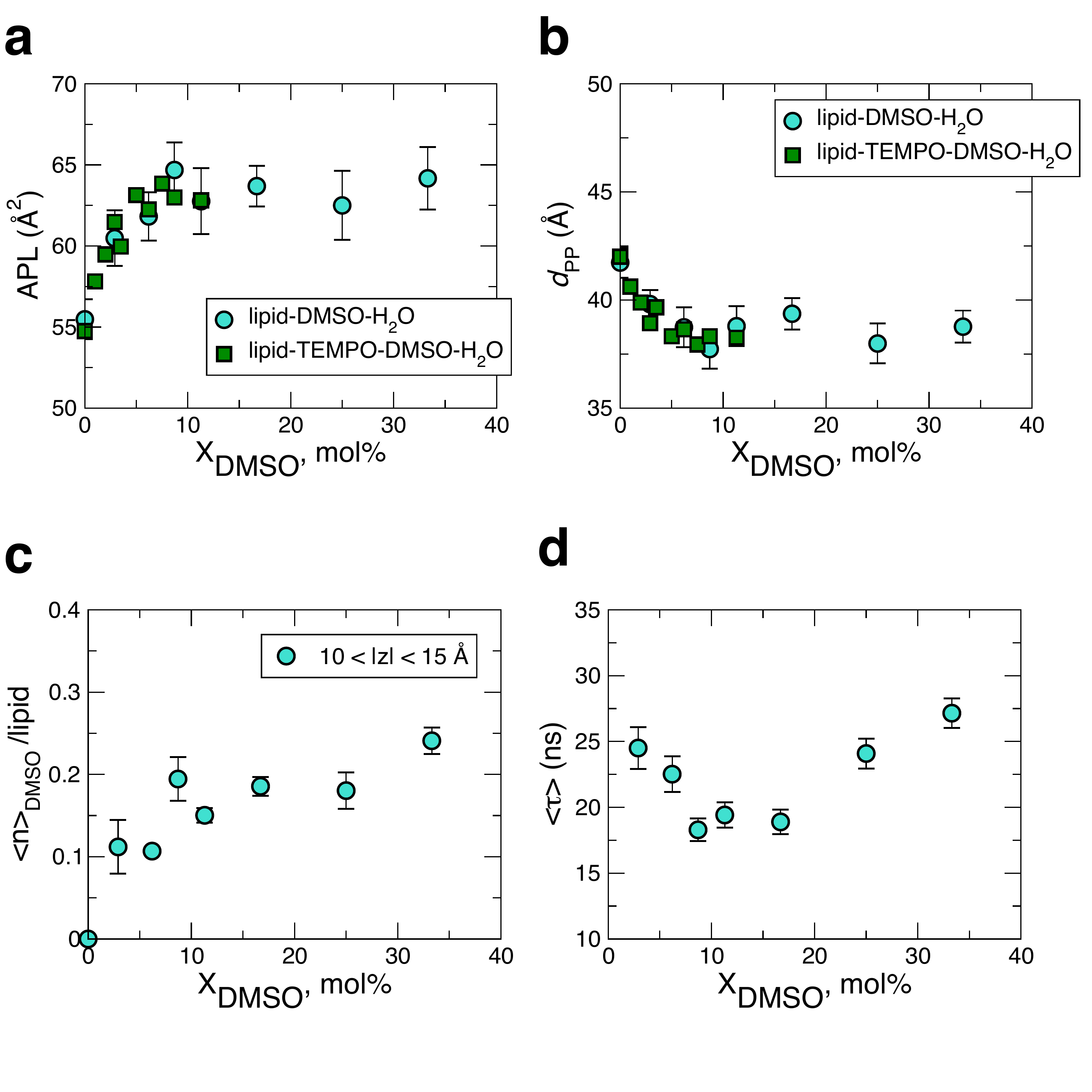}
	\caption{
		Bilayer properties calculated using (a) the area per lipid (APL) and (b) phosphate-phosphate distance ($d_\text{PP}$), at various DMSO  concentrations. The error bars in (a), (b) are the standard deviation.  
		(c) The average number of DMSO molecules trapped around lipid molecule in the range of $10<|z|<15$ \AA\ as a function of $X_{\text{DMSO}}$. 
		(d) The average time for DMSO trapped in $10<|z|<15$ \AA\ to escape beyond $|z|>22$ \AA.  
		\label{APL_and_D_HH}}
\end{figure}

\noindent{\it Surface water diffusion}: 
We have reproduced the experimental observation  \cite{cheng2015BJ} on ``surface water," the increase of surface water diffusion with increasing $X_{\text{DMSO}}$ by explicitly modeling the Tempo and analyzing the water dynamics around it after the Tempo was equilibrated in the interior of bilayer surface. 
By contrast, the water diffusivity calculated \emph{without} Tempo in Fig.\ref{Diffusion}a does not show such a clear increase with $X_{\text{DMSO}}$ in the range of Tempo moiety's equilibrium position ($10\lesssim |z|\lesssim 15$).  
A hint to this conundrum may lie in the dependence of area per lipid (APL) and bilayer thickness ($d_{\text{PP}}$) (Figs.\ref{APL_and_D_HH}a,b). 
In the range of $0<X_{\text{DMSO}}<10$ mol\%, the APL increases from 55 \AA$^2$ to $\sim$ 64 \AA$^2$ (Fig.\ref{APL_and_D_HH}a), whereas $d_{\text{PP}}$ decreases from 41.7 \AA\ to 38.2 \AA\ (Fig.\ref{APL_and_D_HH}b) \cite{yu1998BiophysChem}, a change that could effectively be induced by applying a lateral tension to the bilayer.
 
Although the measurements were done on multilamellar structure of DPPC, electron density profile across bilayer \cite{yu1998BiophysChem} reported a decrease of bilayer thickness due to DMSO. They reported 16 \% decrease of bilayer thickness from 3.79 nm to 3.16 nm in 35 wt\% ($\approx$ 11 mol\%) DMSO after subtracting the contribution of sulfur atom of DMSO trapped in bilayer headgroup from the electron density profile, and the extent of decreases in bilayer thickness is even greater than our result (3.82 nm/4.17 nm $\approx$  8 \%). 
Other studies \cite{kiselev1999dmso}, again on multilamellar structure, argue the constant bilayer thickness up to $X_{\text{DMSO}}\approx 0.3-0.4$, contradicting to the conclusion of the above-mentioned electron density profile study \cite{yu1998BiophysChem}. 
Compared with bilayer stacked in gel phase, a certain amount of cosolvent deposition into bilayer in fluid phase is physically more plausible. 
Thus, it appears that the issue of how DMSO ($X_{\text{DMSO}}<0.1$) affects the bilayer thickness, especially for fluid phase unilamellar structure, remains inconclusive. 
While a possibility of an imperfect force field cannot completely be ruled out, 
our simulation study straightforwardly indicates the increase (decrease) of APL ($d_{\text{PP}}$) of ``unilamellar'' bilayers in the fluid phase (Fig.\ref{APL_and_D_HH}a,b).        

The amount of DMSO deposited below the headgroup ($10<|z|<15$ \AA) increases with $X_{\text{DMSO}}$ (Fig.\ref{APL_and_D_HH}c), providing more free space between lipids, increasing APL; thus contributing to the enhancement of water diffusivity around the Tempo buried below the bilayer surface (Fig.\ref{Dsurf_vs_Dbulk}a).  
The non-monotonic dependence of the escape time of DMSO trapped below the bilayer surface on $X_{\text{DMSO}}$ is also noteworthy (Fig.\ref{APL_and_D_HH}d). 
The first decrease of the mean escape time to the surface ($X_{\text{DMSO}}<10$ \%) may well be an outcome of the increased APL, but the next increase of the mean escape time for $X_{\text{DMSO}}>10$ \% could be related to the increase of $X_{\text{DMSO}}$ while the APL is already saturated (Fig.\ref{APL_and_D_HH}a). While it is not clear how significant the contribution of the increase (decrease) of APL ($d_{\text{PP}}$) with $X_{\text{DMSO}}$ is to the dehydration from the solvent-bilayer interface, 
the increase of APL is certainly a relevant factor that enhances the diffusivity of water in the interior of bilayers. 
\\

\begin{figure}[ht]
\centering
	\includegraphics[width=1.0\columnwidth]{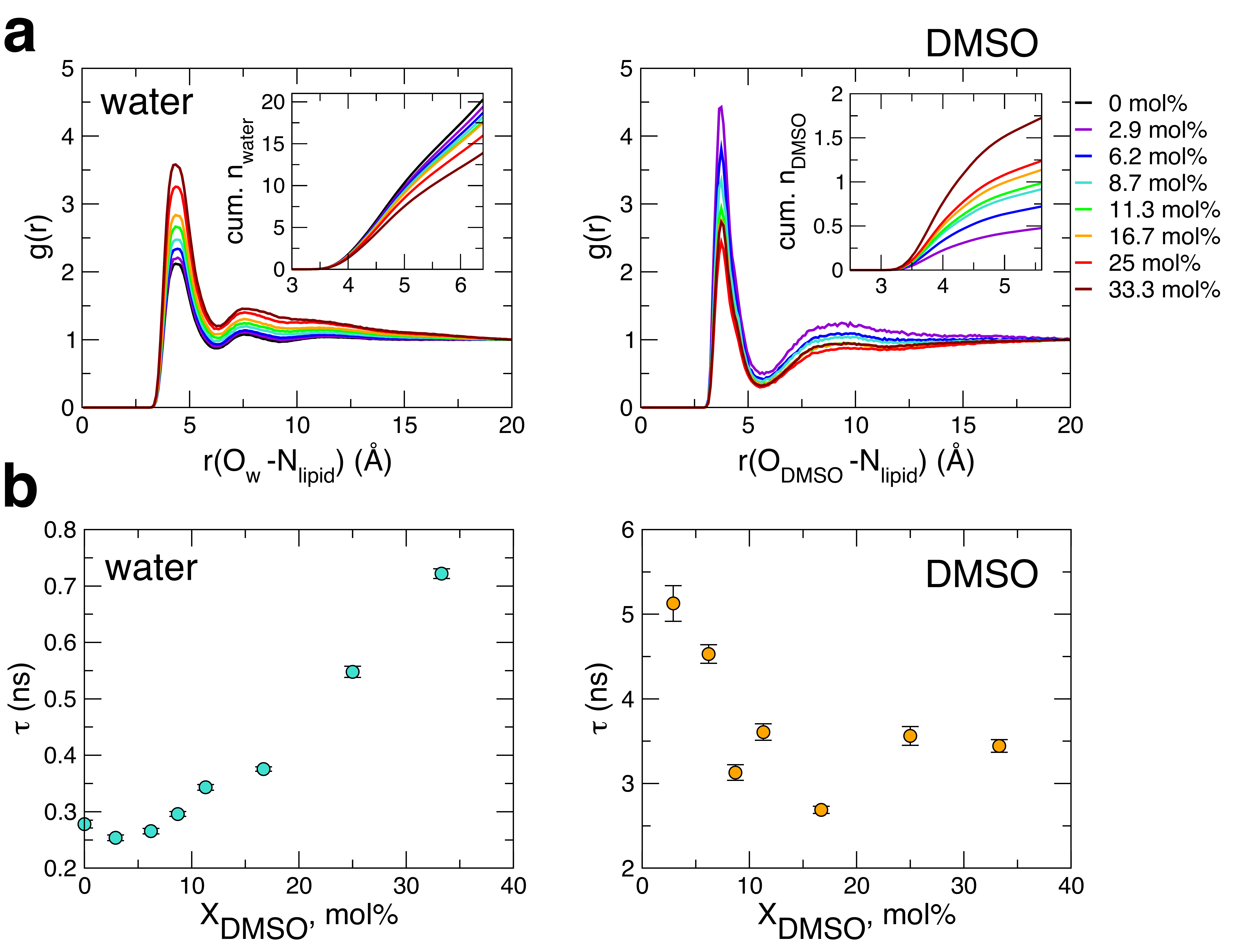}
	\caption{Structure and lifetimes of water and DMSO around choline groups. 
		(a) RDF of water oxygen (left) and DMSO oxygen (right) relative to the nitrogen atom of choline group in varying $X_{\text{DMSO}}$.  
		The insets show the cumulative number of water and DMSO around choline group as a function of $r$. 
		(b) The lifetimes of water (left) and DMSO (right) in the first solvation shell around choline group on bilayer surfaces as a function of $X_{\text{DMSO}}$. 
		\label{Fig_lipid_N_water_DMSO}}
\end{figure}

\noindent{\it Water dynamics around the choline group}:
Based on the result from pulse field gradient (PFG) NMR measurements, decrease of the hydration radii of both DMP$^-$ and TMA$^+$ with increasing DMSO, which is also quantitatively reproduced using our simulations (see SI text and Fig.\ref{FigS_DMP_TMA}d), 
Schrader \emph{et al.} \cite{Schrader2015PNAS} conjectured that DMSO \emph{weakens water binding to PC head-groups}  on bilayer surfaces and thus shortens the range of the repulsive force.  
However, first, the reduction of hydrodynamic radius of DMP$^-$ and TMA$^+$ alone cannot be used for assessing the solvent stability (or lifetime) around DMP$^-$ and TMA$^+$. 
Second, it is not clear whether the measurement in the bulk phase can be used to explain observed phenomena on the bilayer surface where the density of PC head-groups is much higher.  
In stark contrast to their conjecture, we find that both lifetimes of water and DMSO in the first solvation shell around DMP$^-$ and TMA$^+$ in the bulk increase with $X_{\text{DMSO}}$ (see Fig.\ref{FigS_DMP_TMA}e), which indicates that an increased charge-to-size ratio ($e/r_{\text{H}}$) of DMP$^-$ and TMA$^+$ contributes to the stability of the inner solvation shell \cite{Koculi07JACS}. 

In order to gain more microscopic insight, we next examined the solvent structure and its lifetime in the 1st solvation shell around a choline group of the phospholipid bilayer. 
The results summarized in Fig.\ref{Fig_lipid_N_water_DMSO} underscore three points:
(1) the pair correlation between water and the choline group (nitrogen atom) at bilayer surfaces shows an increase  in the 1st solvation shell with increasing $X_{\text{DMSO}}$ (Fig.\ref{Fig_lipid_N_water_DMSO}a); 
(2) the number of water molecules surrounding a PC group decreases with increasing $X_{\text{DMSO}}$ (Fig.\ref{Fig_lipid_N_water_DMSO}a, inset); 
(3) the lifetime of water in the 1st solvation shell around a choline group increases with $X_{\text{DMSO}}$ and is an order of magnitude greater than that around TMA$^+$ in the bulk phase (Fig.\ref{Fig_lipid_N_water_DMSO}b). 
Hence, the presence of DMSO in solution leads to stabilizing water-choline group interaction and increases the lifetime of water. 
This is fully consistent with the decreasing diffusivity of surface water at $|z|\approx 22$ \AA\ with increasing $X_{\text{DMSO}}$.    
\\

\noindent {\bf Concluding Remarks.} 
This study is based on classical MD simulations which disregard quantum mechanical effects, such as polarization and ionization of water molecules near the zwitterionic PC head-group environment. 
It is not clear to what extent these quantum mechanical effects would change our interpretations of DMSO-induced surface water properties.
Nevertheless, the semi-quantitative agreement between experimental measurements and simulation results on the water diffusivity probed with Tempo (see Fig.\ref{Dsurf_vs_Dbulk}) and on the hydration radii of DMP$^-$ and TMA$^+$ (Fig.\ref{FigS_DMP_TMA}d) justify the use of classical MD simulations as a computational tool. 

In conclusion, 
our study shows clearly that DMSO dehydrates surface water from phospholipid bilayers. 
Concerning the more subtle point on the dynamics of water and DMSO interacting with the PC head-group at the solvent-bilayer interface, the DMSO-enhanced surface water diffusion reported by Cheng \emph{et al.} \cite{cheng2015BJ}
is very likely an artifact of Tempo moieties probing the water dynamics at a location below the solvent-bilayer interface.  
A label-free measurement of surface water diffusion (at $|z|\approx 22$ \AA) will reveal that it is a decreasing function of $X_{\text{DMSO}}$ 
(see Fig.\ref{Diffusion}a at $|z|\approx 22$ \AA\ and the data in cyan symbol in Fig.\ref{DMSO_diffusivity}b). 
The spin-label NMR measurement, at present, is the only tool that allows us to directly probe the water  dynamics on biological surfaces, such as proteins and nucleic acids \cite{franck2014JACS,franck2015JACS}. 
Given its significance, the actual position of equilibrated Tempo moiety in lipid bilayers, discussed in this study, and its consequence to the measurement call for a careful re-evaluation of the current biophysical techniques and accompanying theories.   
\\

\section*{Acknowledgements}
We are grateful to Prof. Mahn Won Kim for useful comments, and Alex Schrader and Prof. Songi Han for critical feedback on our work. 
We thank Center for Advanced Computation in KIAS for providing computing resources. 
\\


\clearpage

\section{Supporting Information}
\setcounter{figure}{0}
\makeatletter 
\renewcommand{\thefigure}{S\@arabic\c@figure}
\renewcommand{\theequation}{S\@arabic\c@equation}
\renewcommand{\thetable}{S\@arabic\c@table}
\makeatother 

{\bf Effects of sucrose on surface water. }
In order to assure that DMSO effect on water as a cryoprotectant is specific, we also studied the effect of another cosolvent, sucrose, on surface water. 

First, differences of sucrose from DMSO are clear from the density profiles calculated for water and sucrose. 
The water density starts to deviate from the bulk value farther away from the solvent-bilayer interface (Fig.\ref{sucrose}a), resulting in a greater interface width (greater value of $\xi$. See Fig.\ref{xi_Tetrahedral}a).  
An increase of the cosovlent concentration changes the interface width in opposite direction as compared to DMSO; 
$\xi$ increases with the sucrose concentration ($X_{\text{sucrose}}$), whereas it decreases with $X_{\text{DMSO}}$ (Fig.\ref{xi_Tetrahedral}a). 
Furthermore, in contrast to DMSO solution, a water rich layer is no longer observed in sucrose solution even at high $X_{\text{sucrose}}$ (Fig.\ref{xi_Tetrahedral}a), and sucrose molecules accumulates on the bilayer surface (Fig.\ref{sucrose}b).   
Importantly, $\langle \mathcal{Q}\rangle(z)$ indicates that the tetrahedral structure of the water H-bond network \cite{Errington2001Nature,kumar2009PNAS} is better preserved in sucrose than in DMSO solution (Compare Fig.\ref{sucrose}c with Fig.\ref{Fig_DMSO}d, or see Fig.\ref{xi_Tetrahedral}b),  suggesting that DMSO is a better cryoprotectant.

Next, the local diffusivity profiles of water in the sucrose solution show qualitative difference from those in DMSO solution. 
In the sucrose solution, both $D_{\text{w}}(z;X_{\text{sucrose}})$ and $D_{\text{w}}(z;X_{\text{sucrose}})/D_{\text{w}}^{\text{bulk}}(X_{\text{sucrose}})$ decrease monotonically from the bulk to bilayer, and the water diffusivity hump is no longer observed (Figs.\ref{sucrose}d, e, f). 
Furthermore, unlike DMSO solutions, the diffusivity profiles $D_{\text{w}}(z;X_{\text{sucrose}})/D_{\text{w}}^{\text{bulk}}(X_{\text{sucrose}})$ collapse onto a single curve in sucrose solution (Fig.\ref{sucrose}f). 
The qualitatively different effects of DMSO and sucrose on the surface water dynamics are highlighted by plotting the surface-to-bulk ratio of water diffusion constant, $q_{\text{sucrose}}\equiv D^{\text{surf}}_{\text{w}}/D^{\text{bulk}}_{\text{w}}$ as a function of $X_{\text{sucrose}}$ (Fig.\ref{DMSO_diffusivity}a) where $D_{\text{w}}^{\text{surf}}$ was calculated at $|z|\approx 22$ \AA.  
While $q_{\text{sucrose}}$ does not change with increasing $X_{\text{sucrose}}$, 
an increase of $q_{\text{DMSO}}$ with $X_{\text{DMSO}}$ is evident (Fig.\ref{DMSO_diffusivity}a). 
This suggests that the surface water dynamics is relatively insensitive to DMSO, while both the surface and bulk water dynamics are equally perturbed by sucrose molecules. 
\\

{\bf Hydrodynamic radii of DMP$^-$ and TMA$^+$ in DMSO solution.} 
To corroborate the Schrader \emph{et al.}'s experimental result using pulse field gradient NMR measurement on the hydrodynamic radii of DMP$^-$ and TMA$^+$ in DMSO solution \cite{Schrader2015PNAS} as well as to check the reliability of the molecular force field (Berger force field) used for phospholipids, we obtain the DMSO-dependent hydrodynamic radii ($r_H=k_BT/6\pi\eta D$) of DMP$^-$ and TMA$^+$ in the bulk by calculating both diffusion constant $D$ from the mean square displacements and solution viscosity  ($\eta$) (see Fig.\ref{FigS_DMP_TMA} and its caption for details). 
As shown in Fig.\ref{FigS_DMP_TMA}d, 
excellent agreement is found for $r_H$ for DMP$^-$ and TMA$^+$ between the simulations and PFG NMR measurements. 
Together with the semi-quantitative agreement between simulation and experiment on the bulk/surface water diffusion constants around Tempo (Fig.\ref{Dsurf_vs_Dbulk}) this result gives credence to our simulation results.


\begin{thebibliography}{47}
\expandafter\ifx\csname natexlab\endcsname\relax\def\natexlab#1{#1}\fi
\expandafter\ifx\csname bibnamefont\endcsname\relax
  \def\bibnamefont#1{#1}\fi
\expandafter\ifx\csname bibfnamefont\endcsname\relax
  \def\bibfnamefont#1{#1}\fi
\expandafter\ifx\csname citenamefont\endcsname\relax
  \def\citenamefont#1{#1}\fi
\expandafter\ifx\csname url\endcsname\relax
  \def\url#1{\texttt{#1}}\fi
\expandafter\ifx\csname urlprefix\endcsname\relax\def\urlprefix{URL }\fi
\providecommand{\bibinfo}[2]{#2}
\providecommand{\eprint}[2][]{\url{#2}}

\bibitem[{\citenamefont{Lovelock and Bishop}(1959)}]{Lovelock1959Nature}
\bibinfo{author}{\bibfnamefont{J.}~\bibnamefont{Lovelock}} \bibnamefont{and}
  \bibinfo{author}{\bibfnamefont{M.}~\bibnamefont{Bishop}},
  \bibinfo{journal}{Nature} \textbf{\bibinfo{volume}{183}},
  \bibinfo{pages}{1394} (\bibinfo{year}{1959}).

\bibitem[{\citenamefont{Mazur}(1970)}]{Mazur1970Science}
\bibinfo{author}{\bibfnamefont{P.}~\bibnamefont{Mazur}},
  \bibinfo{journal}{Science} \textbf{\bibinfo{volume}{168}},
  \bibinfo{pages}{939} (\bibinfo{year}{1970}).

\bibitem[{\citenamefont{Anchordoguy et~al.}(1991)\citenamefont{Anchordoguy,
  Cecchini, Crowe, and Crowe}}]{anchordoguy1991Cryobiology}
\bibinfo{author}{\bibfnamefont{T.~J.} \bibnamefont{Anchordoguy}},
  \bibinfo{author}{\bibfnamefont{C.~A.} \bibnamefont{Cecchini}},
  \bibinfo{author}{\bibfnamefont{J.~H.} \bibnamefont{Crowe}}, \bibnamefont{and}
  \bibinfo{author}{\bibfnamefont{L.~M.} \bibnamefont{Crowe}},
  \bibinfo{journal}{Cryobiology} \textbf{\bibinfo{volume}{28}},
  \bibinfo{pages}{467} (\bibinfo{year}{1991}).

\bibitem[{\citenamefont{Soper and Luzar}(1992)}]{soper1992JCP}
\bibinfo{author}{\bibfnamefont{A.}~\bibnamefont{Soper}} \bibnamefont{and}
  \bibinfo{author}{\bibfnamefont{A.}~\bibnamefont{Luzar}}, \bibinfo{journal}{J.
  Chem. Phys.} \textbf{\bibinfo{volume}{97}}, \bibinfo{pages}{1320}
  (\bibinfo{year}{1992}).

\bibitem[{\citenamefont{Luzar and Chandler}(1993)}]{Luzar93JCP}
\bibinfo{author}{\bibfnamefont{A.}~\bibnamefont{Luzar}} \bibnamefont{and}
  \bibinfo{author}{\bibfnamefont{D.}~\bibnamefont{Chandler}},
  \bibinfo{journal}{J. Chem. Phys.} \textbf{\bibinfo{volume}{98}},
  \bibinfo{pages}{8160} (\bibinfo{year}{1993}).

\bibitem[{\citenamefont{Schrader et~al.}(2015)\citenamefont{Schrader,
  Donaldson, Song, Cheng, Lee, Han, and Israelachvili}}]{Schrader2015PNAS}
\bibinfo{author}{\bibfnamefont{A.~M.} \bibnamefont{Schrader}},
  \bibinfo{author}{\bibfnamefont{S.~H.} \bibnamefont{Donaldson}},
  \bibinfo{author}{\bibfnamefont{J.}~\bibnamefont{Song}},
  \bibinfo{author}{\bibfnamefont{C.-Y.} \bibnamefont{Cheng}},
  \bibinfo{author}{\bibfnamefont{D.~W.} \bibnamefont{Lee}},
  \bibinfo{author}{\bibfnamefont{S.}~\bibnamefont{Han}}, \bibnamefont{and}
  \bibinfo{author}{\bibfnamefont{J.~N.} \bibnamefont{Israelachvili}},
  \bibinfo{journal}{Proc. Natl. Acad. Sci. U. S. A.}
  \textbf{\bibinfo{volume}{112}}, \bibinfo{pages}{10708}
  (\bibinfo{year}{2015}).

\bibitem[{\citenamefont{Gordeliy et~al.}(1998)\citenamefont{Gordeliy, Kiselev,
  Lesieur, Pole, and Teixeira}}]{Gordeliy1998BJ}
\bibinfo{author}{\bibfnamefont{V.}~\bibnamefont{Gordeliy}},
  \bibinfo{author}{\bibfnamefont{M.}~\bibnamefont{Kiselev}},
  \bibinfo{author}{\bibfnamefont{P.}~\bibnamefont{Lesieur}},
  \bibinfo{author}{\bibfnamefont{A.}~\bibnamefont{Pole}}, \bibnamefont{and}
  \bibinfo{author}{\bibfnamefont{J.}~\bibnamefont{Teixeira}},
  \bibinfo{journal}{Biophys. J.} \textbf{\bibinfo{volume}{75}},
  \bibinfo{pages}{2343} (\bibinfo{year}{1998}).

\bibitem[{\citenamefont{Kiselev et~al.}(1999)\citenamefont{Kiselev, Lesieur,
  Kisselev, Grabielle-Madelmond, and Ollivon}}]{kiselev1999dmso}
\bibinfo{author}{\bibfnamefont{M.}~\bibnamefont{Kiselev}},
  \bibinfo{author}{\bibfnamefont{P.}~\bibnamefont{Lesieur}},
  \bibinfo{author}{\bibfnamefont{A.}~\bibnamefont{Kisselev}},
  \bibinfo{author}{\bibfnamefont{C.}~\bibnamefont{Grabielle-Madelmond}},
  \bibnamefont{and} \bibinfo{author}{\bibfnamefont{M.}~\bibnamefont{Ollivon}},
  \bibinfo{journal}{J. Alloys Comp.} \textbf{\bibinfo{volume}{286}},
  \bibinfo{pages}{195} (\bibinfo{year}{1999}).

\bibitem[{\citenamefont{Cheng et~al.}(2015)\citenamefont{Cheng, Song, Pas,
  Meijer, and Han}}]{cheng2015BJ}
\bibinfo{author}{\bibfnamefont{C.-Y.} \bibnamefont{Cheng}},
  \bibinfo{author}{\bibfnamefont{J.}~\bibnamefont{Song}},
  \bibinfo{author}{\bibfnamefont{J.}~\bibnamefont{Pas}},
  \bibinfo{author}{\bibfnamefont{L.~H.} \bibnamefont{Meijer}},
  \bibnamefont{and} \bibinfo{author}{\bibfnamefont{S.}~\bibnamefont{Han}},
  \bibinfo{journal}{Biophys. J.} \textbf{\bibinfo{volume}{109}},
  \bibinfo{pages}{330} (\bibinfo{year}{2015}).

\bibitem[{\citenamefont{Bhide and Berkowitz}(2005)}]{bhide2005JCP}
\bibinfo{author}{\bibfnamefont{S.~Y.} \bibnamefont{Bhide}} \bibnamefont{and}
  \bibinfo{author}{\bibfnamefont{M.~L.} \bibnamefont{Berkowitz}},
  \bibinfo{journal}{J. Chem. Phys.} \textbf{\bibinfo{volume}{123}},
  \bibinfo{pages}{224702} (\bibinfo{year}{2005}).

\bibitem[{\citenamefont{Berkowitz and Vacha}(2012)}]{Berkowitz2012ACR}
\bibinfo{author}{\bibfnamefont{M.~L.} \bibnamefont{Berkowitz}}
  \bibnamefont{and} \bibinfo{author}{\bibfnamefont{R.}~\bibnamefont{Vacha}},
  \bibinfo{journal}{Acc. Chem. Res.} \textbf{\bibinfo{volume}{45}},
  \bibinfo{pages}{74} (\bibinfo{year}{2012}).

\bibitem[{\citenamefont{Nieto-Draghi et~al.}(2003)\citenamefont{Nieto-Draghi,
  {\'A}valos, and Rousseau}}]{nieto2003JCP}
\bibinfo{author}{\bibfnamefont{C.}~\bibnamefont{Nieto-Draghi}},
  \bibinfo{author}{\bibfnamefont{J.~B.} \bibnamefont{{\'A}valos}},
  \bibnamefont{and} \bibinfo{author}{\bibfnamefont{B.}~\bibnamefont{Rousseau}},
  \bibinfo{journal}{J. Chem. Phys.} \textbf{\bibinfo{volume}{119}},
  \bibinfo{pages}{4782} (\bibinfo{year}{2003}).

\bibitem[{\citenamefont{Yu and Quinn}(2000)}]{yu2000BBA}
\bibinfo{author}{\bibfnamefont{Z.-W.} \bibnamefont{Yu}} \bibnamefont{and}
  \bibinfo{author}{\bibfnamefont{P.~J.} \bibnamefont{Quinn}},
  \bibinfo{journal}{Biochimica et Biophysica Acta (BBA)-Biomembranes}
  \textbf{\bibinfo{volume}{1509}}, \bibinfo{pages}{440} (\bibinfo{year}{2000}).

\bibitem[{\citenamefont{Berendsen et~al.}(1995)\citenamefont{Berendsen, van~der
  Spoel, and van Drunen}}]{berendsen1995}
\bibinfo{author}{\bibfnamefont{H.~J.} \bibnamefont{Berendsen}},
  \bibinfo{author}{\bibfnamefont{D.}~\bibnamefont{van~der Spoel}},
  \bibnamefont{and} \bibinfo{author}{\bibfnamefont{R.}~\bibnamefont{van
  Drunen}}, \bibinfo{journal}{Comp. Phys. Comm.} \textbf{\bibinfo{volume}{91}},
  \bibinfo{pages}{43} (\bibinfo{year}{1995}).

\bibitem[{\citenamefont{Pronk et~al.}(2013)\citenamefont{Pronk, P{\'a}ll,
  Schulz, Larsson, Bjelkmar, Apostolov, Shirts, Smith, Kasson, van~der Spoel
  et~al.}}]{pronk2013}
\bibinfo{author}{\bibfnamefont{S.}~\bibnamefont{Pronk}},
  \bibinfo{author}{\bibfnamefont{S.}~\bibnamefont{P{\'a}ll}},
  \bibinfo{author}{\bibfnamefont{R.}~\bibnamefont{Schulz}},
  \bibinfo{author}{\bibfnamefont{P.}~\bibnamefont{Larsson}},
  \bibinfo{author}{\bibfnamefont{P.}~\bibnamefont{Bjelkmar}},
  \bibinfo{author}{\bibfnamefont{R.}~\bibnamefont{Apostolov}},
  \bibinfo{author}{\bibfnamefont{M.~R.} \bibnamefont{Shirts}},
  \bibinfo{author}{\bibfnamefont{J.~C.} \bibnamefont{Smith}},
  \bibinfo{author}{\bibfnamefont{P.~M.} \bibnamefont{Kasson}},
  \bibinfo{author}{\bibfnamefont{D.}~\bibnamefont{van~der Spoel}},
  \bibnamefont{et~al.}, \bibinfo{journal}{Bioinformatics}
  \textbf{\bibinfo{volume}{29}}, \bibinfo{pages}{845} (\bibinfo{year}{2013}).

\bibitem[{\citenamefont{Piggot et~al.}(2012)\citenamefont{Piggot, Pi{\~n}eiro,
  and Khalid}}]{piggot2012}
\bibinfo{author}{\bibfnamefont{T.~J.} \bibnamefont{Piggot}},
  \bibinfo{author}{\bibfnamefont{A.}~\bibnamefont{Pi{\~n}eiro}},
  \bibnamefont{and} \bibinfo{author}{\bibfnamefont{S.}~\bibnamefont{Khalid}},
  \bibinfo{journal}{J. Chem. Theor. Comp.} \textbf{\bibinfo{volume}{8}},
  \bibinfo{pages}{4593} (\bibinfo{year}{2012}).

\bibitem[{\citenamefont{Berger et~al.}(1997)\citenamefont{Berger, Edholm, and
  J{\"a}hnig}}]{berger1997}
\bibinfo{author}{\bibfnamefont{O.}~\bibnamefont{Berger}},
  \bibinfo{author}{\bibfnamefont{O.}~\bibnamefont{Edholm}}, \bibnamefont{and}
  \bibinfo{author}{\bibfnamefont{F.}~\bibnamefont{J{\"a}hnig}},
  \bibinfo{journal}{Biophys. J.} \textbf{\bibinfo{volume}{72}},
  \bibinfo{pages}{2002} (\bibinfo{year}{1997}).

\bibitem[{\citenamefont{Hermans et~al.}(1984)\citenamefont{Hermans, Berendsen,
  Van~Gunsteren, and Postma}}]{hermans1984}
\bibinfo{author}{\bibfnamefont{J.}~\bibnamefont{Hermans}},
  \bibinfo{author}{\bibfnamefont{H.~J.} \bibnamefont{Berendsen}},
  \bibinfo{author}{\bibfnamefont{W.~F.} \bibnamefont{Van~Gunsteren}},
  \bibnamefont{and} \bibinfo{author}{\bibfnamefont{J.~P.}
  \bibnamefont{Postma}}, \bibinfo{journal}{Biopolymers}
  \textbf{\bibinfo{volume}{23}}, \bibinfo{pages}{1513} (\bibinfo{year}{1984}).

\bibitem[{\citenamefont{Geerke et~al.}(2004)\citenamefont{Geerke, Oostenbrink,
  van~der Vegt, and van Gunsteren}}]{geerke2004}
\bibinfo{author}{\bibfnamefont{D.~P.} \bibnamefont{Geerke}},
  \bibinfo{author}{\bibfnamefont{C.}~\bibnamefont{Oostenbrink}},
  \bibinfo{author}{\bibfnamefont{N.~F.} \bibnamefont{van~der Vegt}},
  \bibnamefont{and} \bibinfo{author}{\bibfnamefont{W.~F.} \bibnamefont{van
  Gunsteren}}, \bibinfo{journal}{J. Phys. Chem. B}
  \textbf{\bibinfo{volume}{108}}, \bibinfo{pages}{1436} (\bibinfo{year}{2004}).

\bibitem[{\citenamefont{Oostenbrink et~al.}(2004)\citenamefont{Oostenbrink,
  Villa, Mark, and Van~Gunsteren}}]{oostenbrink04JCC}
\bibinfo{author}{\bibfnamefont{C.}~\bibnamefont{Oostenbrink}},
  \bibinfo{author}{\bibfnamefont{A.}~\bibnamefont{Villa}},
  \bibinfo{author}{\bibfnamefont{A.}~\bibnamefont{Mark}}, \bibnamefont{and}
  \bibinfo{author}{\bibfnamefont{W.}~\bibnamefont{Van~Gunsteren}},
  \bibinfo{journal}{J. Comp. Chem.} \textbf{\bibinfo{volume}{25}},
  \bibinfo{pages}{1656} (\bibinfo{year}{2004}).

\bibitem[{\citenamefont{Kyrychenko et~al.}(2014)\citenamefont{Kyrychenko,
  Rodnin, and Ladokhin}}]{kyrychenko2014}
\bibinfo{author}{\bibfnamefont{A.}~\bibnamefont{Kyrychenko}},
  \bibinfo{author}{\bibfnamefont{M.~V.} \bibnamefont{Rodnin}},
  \bibnamefont{and} \bibinfo{author}{\bibfnamefont{A.~S.}
  \bibnamefont{Ladokhin}}, \bibinfo{journal}{J. Membr. Biol.} pp.
  \bibinfo{pages}{1--12} (\bibinfo{year}{2014}).

\bibitem[{\citenamefont{Errington and Debenedetti}(2001)}]{Errington2001Nature}
\bibinfo{author}{\bibfnamefont{J.~R.} \bibnamefont{Errington}}
  \bibnamefont{and} \bibinfo{author}{\bibfnamefont{P.~G.}
  \bibnamefont{Debenedetti}}, \bibinfo{journal}{Nature}
  \textbf{\bibinfo{volume}{409}}, \bibinfo{pages}{318} (\bibinfo{year}{2001}).

\bibitem[{\citenamefont{Kumar et~al.}(2009)\citenamefont{Kumar, Buldyrev, and
  Stanley}}]{kumar2009PNAS}
\bibinfo{author}{\bibfnamefont{P.}~\bibnamefont{Kumar}},
  \bibinfo{author}{\bibfnamefont{S.~V.} \bibnamefont{Buldyrev}},
  \bibnamefont{and} \bibinfo{author}{\bibfnamefont{H.~E.}
  \bibnamefont{Stanley}}, \bibinfo{journal}{Proc. Natl. Acad. Sci. U. S. A.}
  \textbf{\bibinfo{volume}{106}}, \bibinfo{pages}{22130}
  (\bibinfo{year}{2009}).

\bibitem[{\citenamefont{Lounnas et~al.}(1994)\citenamefont{Lounnas, Pettitt,
  and Phillips~Jr}}]{lounnas1994BJ}
\bibinfo{author}{\bibfnamefont{V.}~\bibnamefont{Lounnas}},
  \bibinfo{author}{\bibfnamefont{B.}~\bibnamefont{Pettitt}}, \bibnamefont{and}
  \bibinfo{author}{\bibfnamefont{G.}~\bibnamefont{Phillips~Jr}},
  \bibinfo{journal}{Biophys. J.} \textbf{\bibinfo{volume}{66}},
  \bibinfo{pages}{601} (\bibinfo{year}{1994}).

\bibitem[{\citenamefont{Rahman and Stillinger}(1971)}]{Rahman1971JCP}
\bibinfo{author}{\bibfnamefont{A.}~\bibnamefont{Rahman}} \bibnamefont{and}
  \bibinfo{author}{\bibfnamefont{F.~H.} \bibnamefont{Stillinger}},
  \bibinfo{journal}{J. Chem. Phys.} \textbf{\bibinfo{volume}{55}},
  \bibinfo{pages}{3336} (\bibinfo{year}{1971}).

\bibitem[{\citenamefont{Mills}(1973)}]{mills1973JPC}
\bibinfo{author}{\bibfnamefont{R.}~\bibnamefont{Mills}}, \bibinfo{journal}{J.
  Phys. Chem.} \textbf{\bibinfo{volume}{77}}, \bibinfo{pages}{685}
  (\bibinfo{year}{1973}).

\bibitem[{\citenamefont{Mark and Nilsson}(2001)}]{Mark02JPCA}
\bibinfo{author}{\bibfnamefont{P.}~\bibnamefont{Mark}} \bibnamefont{and}
  \bibinfo{author}{\bibfnamefont{L.}~\bibnamefont{Nilsson}},
  \bibinfo{journal}{J. Phys. Chem. A} \textbf{\bibinfo{volume}{105}},
  \bibinfo{pages}{9954} (\bibinfo{year}{2001}).

\bibitem[{\citenamefont{Armstrong and Han}(2009)}]{armstrong2009JACS}
\bibinfo{author}{\bibfnamefont{B.~D.} \bibnamefont{Armstrong}}
  \bibnamefont{and} \bibinfo{author}{\bibfnamefont{S.}~\bibnamefont{Han}},
  \bibinfo{journal}{J. Am. Chem. Soc.} \textbf{\bibinfo{volume}{131}},
  \bibinfo{pages}{4641} (\bibinfo{year}{2009}).

\bibitem[{\citenamefont{Smondyrev and Berkowitz}(1999)}]{smondyrev1999BJ}
\bibinfo{author}{\bibfnamefont{A.~M.} \bibnamefont{Smondyrev}}
  \bibnamefont{and} \bibinfo{author}{\bibfnamefont{M.~L.}
  \bibnamefont{Berkowitz}}, \bibinfo{journal}{Biophys. J.}
  \textbf{\bibinfo{volume}{77}}, \bibinfo{pages}{2075} (\bibinfo{year}{1999}).

\bibitem[{\citenamefont{Hughes et~al.}(2012)\citenamefont{Hughes, Mark, and
  Mancera}}]{hughes2012JPCB}
\bibinfo{author}{\bibfnamefont{Z.~E.} \bibnamefont{Hughes}},
  \bibinfo{author}{\bibfnamefont{A.~E.} \bibnamefont{Mark}}, \bibnamefont{and}
  \bibinfo{author}{\bibfnamefont{R.~L.} \bibnamefont{Mancera}},
  \bibinfo{journal}{J. Phys. Chem. B} \textbf{\bibinfo{volume}{116}},
  \bibinfo{pages}{11911} (\bibinfo{year}{2012}).

\bibitem[{\citenamefont{Sum and de~Pablo}(2003)}]{Sum2003BJ}
\bibinfo{author}{\bibfnamefont{A.~K.} \bibnamefont{Sum}} \bibnamefont{and}
  \bibinfo{author}{\bibfnamefont{J.~J.} \bibnamefont{de~Pablo}},
  \bibinfo{journal}{Biophys. J.} \textbf{\bibinfo{volume}{85}},
  \bibinfo{pages}{3636} (\bibinfo{year}{2003}).

\bibitem[{\citenamefont{Notman et~al.}(2006)\citenamefont{Notman, Noro,
  O'Malley, and Anwar}}]{notman2006JACS}
\bibinfo{author}{\bibfnamefont{R.}~\bibnamefont{Notman}},
  \bibinfo{author}{\bibfnamefont{M.}~\bibnamefont{Noro}},
  \bibinfo{author}{\bibfnamefont{B.}~\bibnamefont{O'Malley}}, \bibnamefont{and}
  \bibinfo{author}{\bibfnamefont{J.}~\bibnamefont{Anwar}}, \bibinfo{journal}{J.
  Am. Chem. Soc.} \textbf{\bibinfo{volume}{128}}, \bibinfo{pages}{13982}
  (\bibinfo{year}{2006}).

\bibitem[{\citenamefont{Gurtovenko and Anwar}(2007)}]{gurtovenko2007JPCB}
\bibinfo{author}{\bibfnamefont{A.~A.} \bibnamefont{Gurtovenko}}
  \bibnamefont{and} \bibinfo{author}{\bibfnamefont{J.}~\bibnamefont{Anwar}},
  \bibinfo{journal}{J. Phys. Chem. B} \textbf{\bibinfo{volume}{111}},
  \bibinfo{pages}{10453} (\bibinfo{year}{2007}).

\bibitem[{\citenamefont{Cahn and Hilliard}(1958)}]{Cahn1958JCP}
\bibinfo{author}{\bibfnamefont{J.~W.} \bibnamefont{Cahn}} \bibnamefont{and}
  \bibinfo{author}{\bibfnamefont{J.~E.} \bibnamefont{Hilliard}},
  \bibinfo{journal}{J. Chem. Phys.} \textbf{\bibinfo{volume}{28}},
  \bibinfo{pages}{258} (\bibinfo{year}{1958}).

\bibitem[{\citenamefont{Bu et~al.}(2014)\citenamefont{Bu, Kim, and
  Vaknin}}]{bu2014JPCC}
\bibinfo{author}{\bibfnamefont{W.}~\bibnamefont{Bu}},
  \bibinfo{author}{\bibfnamefont{D.}~\bibnamefont{Kim}}, \bibnamefont{and}
  \bibinfo{author}{\bibfnamefont{D.}~\bibnamefont{Vaknin}},
  \bibinfo{journal}{J. Phys. Chem. C} \textbf{\bibinfo{volume}{118}},
  \bibinfo{pages}{12405} (\bibinfo{year}{2014}).

\bibitem[{\citenamefont{Subczynski et~al.}(2010)\citenamefont{Subczynski,
  Raguz, and Widomska}}]{Widomska2010MMB}
\bibinfo{author}{\bibfnamefont{W.~K.} \bibnamefont{Subczynski}},
  \bibinfo{author}{\bibfnamefont{M.}~\bibnamefont{Raguz}}, \bibnamefont{and}
  \bibinfo{author}{\bibfnamefont{J.}~\bibnamefont{Widomska}},
  \bibinfo{journal}{Methods Mol. Biol.} \textbf{\bibinfo{volume}{606}},
  \bibinfo{pages}{247} (\bibinfo{year}{2010}).

\bibitem[{\citenamefont{Kyrychenko and Ladokhin}(2013)}]{kyrychenko2013JPCB}
\bibinfo{author}{\bibfnamefont{A.}~\bibnamefont{Kyrychenko}} \bibnamefont{and}
  \bibinfo{author}{\bibfnamefont{A.~S.} \bibnamefont{Ladokhin}},
  \bibinfo{journal}{J. Phys. Chem. B} \textbf{\bibinfo{volume}{117}},
  \bibinfo{pages}{5875} (\bibinfo{year}{2013}).

\bibitem[{\citenamefont{Kyrychenko and Ladokhin}(2014)}]{Ladokhin14AB}
\bibinfo{author}{\bibfnamefont{A.}~\bibnamefont{Kyrychenko}} \bibnamefont{and}
  \bibinfo{author}{\bibfnamefont{A.~S.} \bibnamefont{Ladokhin}},
  \bibinfo{journal}{Anal. Biochem.} \textbf{\bibinfo{volume}{446}},
  \bibinfo{pages}{19} (\bibinfo{year}{2014}).

\bibitem[{\citenamefont{Yu and Quinn}(1998)}]{yu1998BiophysChem}
\bibinfo{author}{\bibfnamefont{Z.-W.} \bibnamefont{Yu}} \bibnamefont{and}
  \bibinfo{author}{\bibfnamefont{P.~J.} \bibnamefont{Quinn}},
  \bibinfo{journal}{Biophys. Chem.} \textbf{\bibinfo{volume}{70}},
  \bibinfo{pages}{35} (\bibinfo{year}{1998}).

\bibitem[{\citenamefont{Koculi et~al.}(2007)\citenamefont{Koculi, Hyeon,
  Thirumalai, and Woodson}}]{Koculi07JACS}
\bibinfo{author}{\bibfnamefont{E.}~\bibnamefont{Koculi}},
  \bibinfo{author}{\bibfnamefont{C.}~\bibnamefont{Hyeon}},
  \bibinfo{author}{\bibfnamefont{D.}~\bibnamefont{Thirumalai}},
  \bibnamefont{and} \bibinfo{author}{\bibfnamefont{S.~A.}
  \bibnamefont{Woodson}}, \bibinfo{journal}{J. Am. Chem. Soc.}
  \textbf{\bibinfo{volume}{129}}, \bibinfo{pages}{2676} (\bibinfo{year}{2007}).

\bibitem[{\citenamefont{Franck et~al.}(2014)\citenamefont{Franck, Sokolovski,
  Kessler, Matalon, Gordon-Grossman, Han, Goldfarb, and
  Horovitz}}]{franck2014JACS}
\bibinfo{author}{\bibfnamefont{J.~M.} \bibnamefont{Franck}},
  \bibinfo{author}{\bibfnamefont{M.}~\bibnamefont{Sokolovski}},
  \bibinfo{author}{\bibfnamefont{N.}~\bibnamefont{Kessler}},
  \bibinfo{author}{\bibfnamefont{E.}~\bibnamefont{Matalon}},
  \bibinfo{author}{\bibfnamefont{M.}~\bibnamefont{Gordon-Grossman}},
  \bibinfo{author}{\bibfnamefont{S.-I.} \bibnamefont{Han}},
  \bibinfo{author}{\bibfnamefont{D.}~\bibnamefont{Goldfarb}}, \bibnamefont{and}
  \bibinfo{author}{\bibfnamefont{A.}~\bibnamefont{Horovitz}},
  \bibinfo{journal}{J. Am. Chem. Soc.} \textbf{\bibinfo{volume}{136}},
  \bibinfo{pages}{9396} (\bibinfo{year}{2014}).

\bibitem[{\citenamefont{Franck et~al.}(2015)\citenamefont{Franck, Ding, Stone,
  Qin, and Han}}]{franck2015JACS}
\bibinfo{author}{\bibfnamefont{J.~M.} \bibnamefont{Franck}},
  \bibinfo{author}{\bibfnamefont{Y.}~\bibnamefont{Ding}},
  \bibinfo{author}{\bibfnamefont{K.}~\bibnamefont{Stone}},
  \bibinfo{author}{\bibfnamefont{P.~Z.} \bibnamefont{Qin}}, \bibnamefont{and}
  \bibinfo{author}{\bibfnamefont{S.}~\bibnamefont{Han}}, \bibinfo{journal}{J.
  Am. Chem. Soc.} \textbf{\bibinfo{volume}{137}}, \bibinfo{pages}{12013}
  (\bibinfo{year}{2015}).

\bibitem[{\citenamefont{Packer and Tomlinson}(1971)}]{packer1971TFS}
\bibinfo{author}{\bibfnamefont{K.}~\bibnamefont{Packer}} \bibnamefont{and}
  \bibinfo{author}{\bibfnamefont{D.}~\bibnamefont{Tomlinson}},
  \bibinfo{journal}{Transactions of the Faraday Society}
  \textbf{\bibinfo{volume}{67}}, \bibinfo{pages}{1302} (\bibinfo{year}{1971}).

\bibitem[{\citenamefont{Borin and Skaf}(1999)}]{Borin1999JCP}
\bibinfo{author}{\bibfnamefont{I.~A.} \bibnamefont{Borin}} \bibnamefont{and}
  \bibinfo{author}{\bibfnamefont{M.~S.} \bibnamefont{Skaf}},
  \bibinfo{journal}{J. Chem. Phys.} \textbf{\bibinfo{volume}{110}},
  \bibinfo{pages}{6412} (\bibinfo{year}{1999}).

\bibitem[{\citenamefont{Price et~al.}(2001)\citenamefont{Price, Ostrovsky, and
  Jorgensen}}]{price2001JCC}
\bibinfo{author}{\bibfnamefont{M.~L.} \bibnamefont{Price}},
  \bibinfo{author}{\bibfnamefont{D.}~\bibnamefont{Ostrovsky}},
  \bibnamefont{and} \bibinfo{author}{\bibfnamefont{W.~L.}
  \bibnamefont{Jorgensen}}, \bibinfo{journal}{J. Comp. Chem.}
  \textbf{\bibinfo{volume}{22}}, \bibinfo{pages}{1340} (\bibinfo{year}{2001}).

\bibitem[{\citenamefont{MacKerell~Jr et~al.}(1998)\citenamefont{MacKerell~Jr,
  Bashford, Bellott, Dunbrack~Jr, Evanseck, Field, Fischer, Gao, Guo, Ha
  et~al.}}]{mackerell98JPCB}
\bibinfo{author}{\bibfnamefont{A.}~\bibnamefont{MacKerell~Jr}},
  \bibinfo{author}{\bibfnamefont{D.}~\bibnamefont{Bashford}},
  \bibinfo{author}{\bibfnamefont{M.}~\bibnamefont{Bellott}},
  \bibinfo{author}{\bibfnamefont{R.}~\bibnamefont{Dunbrack~Jr}},
  \bibinfo{author}{\bibfnamefont{J.}~\bibnamefont{Evanseck}},
  \bibinfo{author}{\bibfnamefont{M.}~\bibnamefont{Field}},
  \bibinfo{author}{\bibfnamefont{S.}~\bibnamefont{Fischer}},
  \bibinfo{author}{\bibfnamefont{J.}~\bibnamefont{Gao}},
  \bibinfo{author}{\bibfnamefont{H.}~\bibnamefont{Guo}},
  \bibinfo{author}{\bibfnamefont{S.}~\bibnamefont{Ha}}, \bibnamefont{et~al.},
  \bibinfo{journal}{J. Phys. Chem. B} \textbf{\bibinfo{volume}{102}},
  \bibinfo{pages}{3586} (\bibinfo{year}{1998}).

\bibitem[{\citenamefont{Palmer}(1994)}]{palmer1994PRE}
\bibinfo{author}{\bibfnamefont{B.~J.} \bibnamefont{Palmer}},
  \bibinfo{journal}{Phys. Rev. E} \textbf{\bibinfo{volume}{49}},
  \bibinfo{pages}{359} (\bibinfo{year}{1994}).

\end{thebibliography}

\begin{table}[ht]
	\caption{\label{table}The area per lipid (APL) and the mean bilayer thickness ($d_\text{PP}$) at various DMSO, sucrose mol\%.}
	\centering 
	\scriptsize
	\begin{tabular*}{0.5\textwidth}
		{@{\extracolsep{\fill}}lccccc}
		\hline\hline
		system & mol\%  & wt\% & Time ($\mu$s) &   APL/\AA$^2$ & $d_\text{PP}$/\AA \\
		\hline
		&  & &  & & \\ 
		lipid- \\DMSO-H$_2$O\textsuperscript{\emph{a}}& 0 & 0 & 0.4 & $55.5\pm 1.2 $ & $41.7\pm 0.7 $ \\ 
		& 2.9 & 11.4 & 0.4 & $60.5\pm 1.7 $ & $39.8\pm 0.7 $ \\ 
		& 6.2 & 22.3 & 0.4 & $61.8\pm 1.5 $ & $38.7\pm 0.9 $ \\ 
		& 8.7 & 29.4 & 0.4 & $64.7\pm 1.7 $ & $37.7\pm 0.9 $ \\ 
		& 11.3 & 35.5 & 0.4 & $62.8\pm 2.0 $ & $38.8\pm 0.9 $ \\ 
		& 16.7 & 46.5 & 0.4 & $63.7\pm 1.3 $ & $39.4\pm 0.7 $ \\ 
		& 25.0 & 59.1 & 0.4 & $62.5\pm 2.1 $ & $38.0\pm 0.9 $ \\ 
		& 33.3 & 68.5 & 0.4 & $64.2\pm 1.9 $ & $38.8\pm 0.7 $ \\ 
		
		lipid- \\sucrose-H$_2$O\textsuperscript{\emph{b}}& 1.5 & 22.3 & 1 & $63.7\pm 2.5 $ & $37.7\pm 1.1 $ \\  
		& 2.8 & 35.5 & 1 & $65.0\pm 3.4 $ & $37.1\pm 1.5 $ \\ 
		& 4.4 & 46.5 & 1 & $62.1\pm 1.6 $ & $38.3\pm 0.9 $ \\ 
		
		lipid-Tempo- \\DMSO-H$_2$O\textsuperscript{\emph{b}}& 0 & 0 & 1 & $54.8\pm 1.0 $ & $42.0\pm 0.7 $ \\ 
		& 1 & 4.2 & 1 & $57.8\pm 1.3 $ & $40.6\pm 0.8 $ \\ 
		& 2 & 8.2 & 1 & $59.5\pm 1.3 $ & $39.9\pm 0.8 $ \\ 
		& 2.9 & 11.4 & 1 & $61.5\pm 1.6 $ & $38.9\pm 0.9 $ \\ 
		& 3.5 & 13.6 & 1 &  $60.0\pm 1.3 $ & $39.7\pm 0.8 $ \\ 
		& 5 & 18.7 & 1 & $63.1\pm 2.0 $ & $38.3\pm 0.9 $ \\ 
		& 6.2 & 22.3 & 1 & $62.3\pm 1.4 $ & $38.6\pm 0.8 $ \\ 
		& 7.5 & 26.0 & 1 & $63.9\pm 1.8 $ & $37.9\pm 0.9 $ \\ 
		& 8.7 & 29.4 & 1 & $63.0\pm 1.6 $ & $38.3\pm 0.8 $ \\ 
		& 11.3 & 35.5 & 1 & $62.8\pm 1.9 $  & $38.3\pm 1.0 $ \\ [1ex]
		\hline\hline& & & & 
	\end{tabular*}
	\label{table:table_systems}
	\begin{flushleft}
		The analysis was done for the last \textsuperscript{\emph{a}} 0.3 and 
		\textsuperscript{\emph{b}} 0.9 $\mu$s. 
		The average positions of phosphorus atom in the upper and lower leaflets were used to calculate the mean thickness of bilayer, $d_\text{PP}$.
	\end{flushleft} 
\end{table} 
\clearpage 

\begin{figure*}
\centering
	\includegraphics[width=1.3\columnwidth]{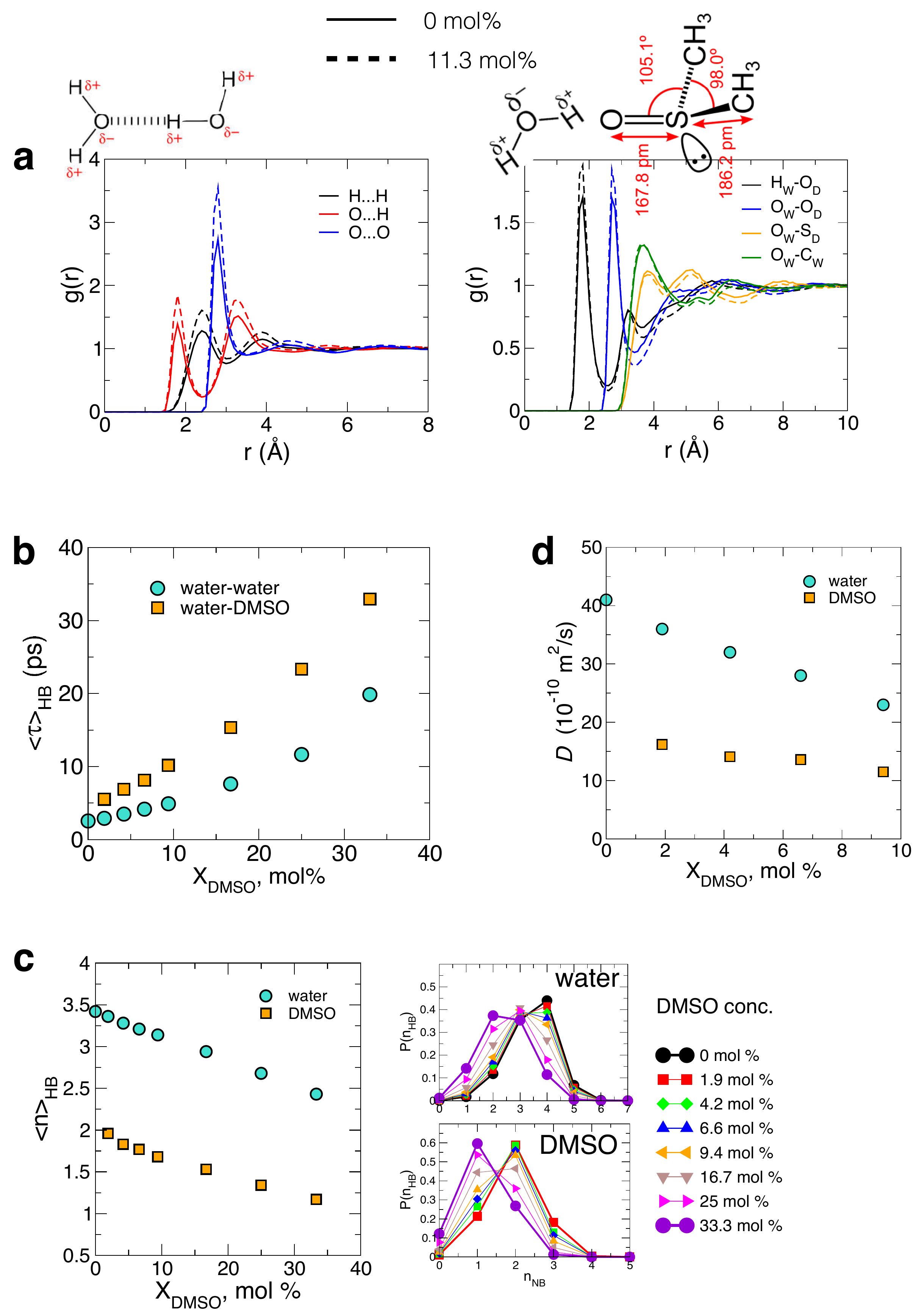}
	\caption{
	Structural and dynamical characteristics of water and DMSO in varying concentrations of DMSO solution: 
	DMSO, an aprotic cosolvent whose dipole moment (3.96 Debye) is greater than that of water (1.85 Debye), 
is H-bond acceptor, capable of forming ``two'' H-bonds with water molecules via sulfonyl group (Fig.\ref{bulk_water_dynamics}c), 
but is not a H-bond donor. 
The presence of DMSO in aqueous solution increases both the water/water and water/DMSO nearest neighbor correlations (Fig.\ref{bulk_water_dynamics}a) and binds more strongly with water than water themselves. 
Thus, DMSO not only increases the H-bond lifetime (Fig.\ref{bulk_water_dynamics}b) but also decreases of the number of H-bonds (Fig.\ref{bulk_water_dynamics}c) and diffusion coefficient of both water and DMSO (Fig.\ref{bulk_water_dynamics}d), the trend of which continues until the concentration of DMSO reaches 33.3 mol\%, at which the 2:1 stoichiometric ratio of water-DMSO is satisfied \cite{Luzar93JCP,packer1971TFS,Borin1999JCP}.  
DMSO disrupts the ``water structure'' beyond the range of nearest molecular neighbors, preventing ice formation at low temperature \cite{Luzar93JCP}.  
To recapitulate, DMSO slows down water dynamics and disrupt the tetrahedral ordering of water structure. 
	[(a) Water-water (left) and water-DMSO (right) radial distribution functions at 0 mol\% and 11.3 mol\% of DMSO solution. 
	DMSO increases inter-molecular correlations. 
	(b) Water-water and water-DMSO H-bond lifetimes as a function of DMSO concentration. 
	(c) Average number of H-bonds ($\langle n_{\text{HB}}\rangle$) around water and DMSO as a function of DMSO concentration (left) and its distribution, $P(n_{\text{HB}})$ for water (top) and DMSO (bottom).  
	(d) Local diffusivity of water and DMSO molecules as a function of DMSO concentration.]
		\label{bulk_water_dynamics}} 
\end{figure*}

\begin{figure}
\centering
	\includegraphics[width=0.9\columnwidth]{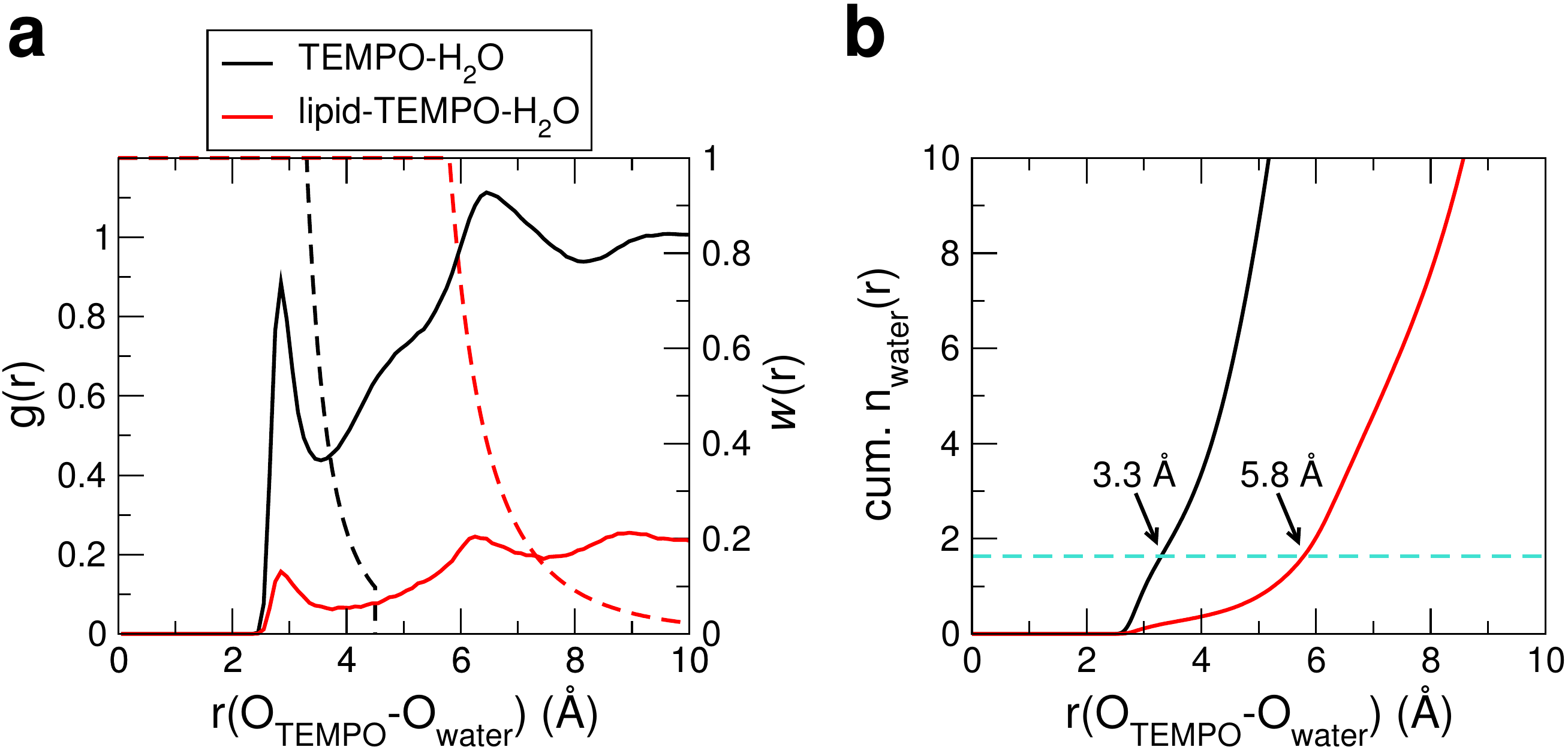}
	\caption{
		Water radial distribution ($g(r)$) around Tempo in bulk (black) and Tempo in bilayer (red). The initial position-dependent weighting factor $w(r)$ for each case is shown in dashed line (axis label on the right). 
		Cumulative number of water molecules up to $r$, $n(r)=4\pi\int_0^rg(r')r'^2dr'$.  
		For Tempo in bulk, $\sigma_{\text{bulk}}=3.3$ \AA\ is the position of the 1st solvation shell. 
		For Tempo in bilayer, we chose $\sigma_{\text{bilayer}}=5.8$ \AA, such that the number of waters probed by the Tempo in bilayer for the lifetime calculation is identical to the number of water probed by the Tempo in the bulk, i.e., $n(\sigma_{\text{bulk}})=n(\sigma_{\text{bilayer}})$.  
			\label{RDF_Tempo_O_vs_water_O}}
\end{figure}

\begin{figure}
\centering
	\includegraphics[width=0.9\columnwidth]{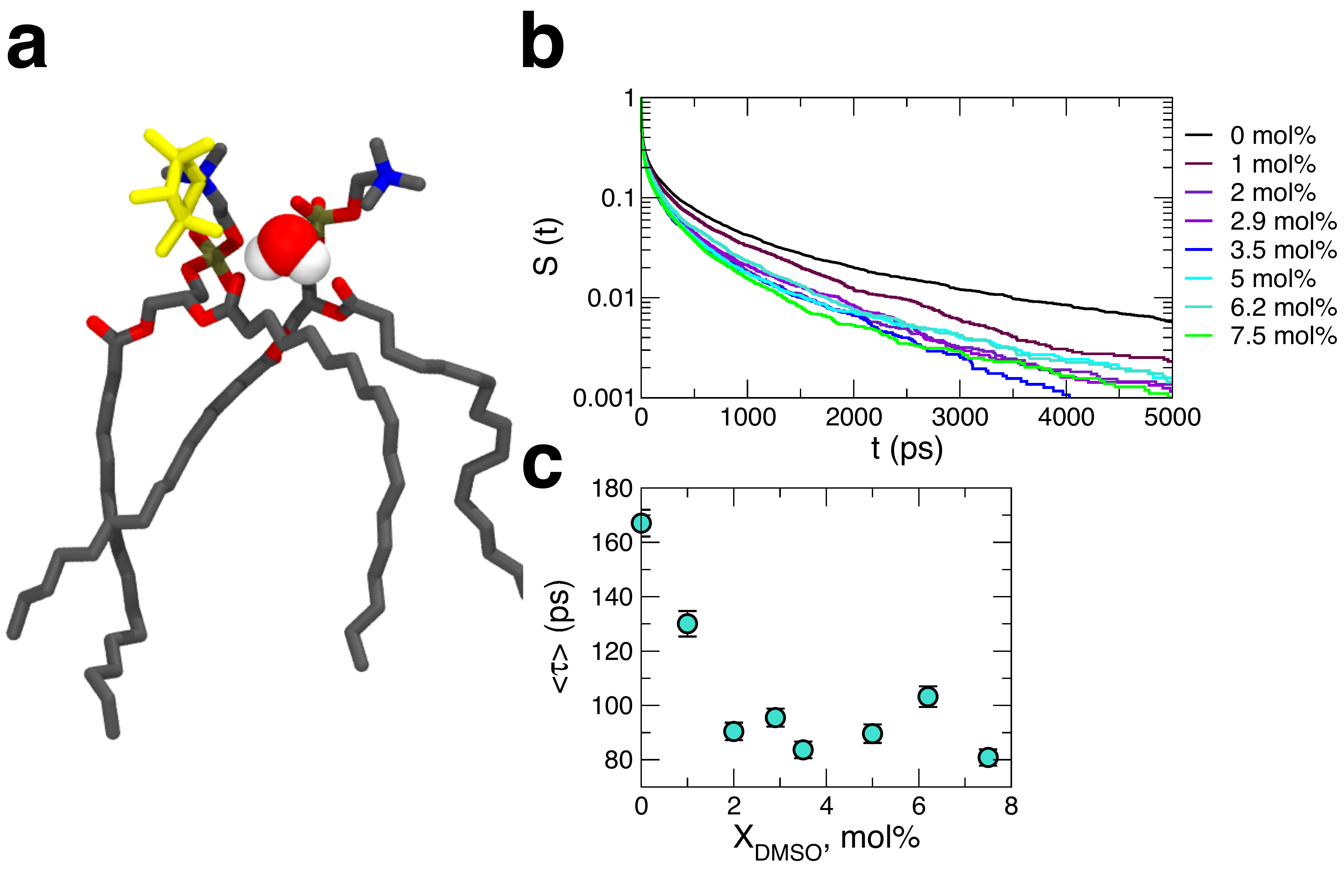}
	\caption{Releasing kinetics of water from Tempo moiety (depicted in yellow). 
		(a) A snapshot of a water molecule trapped between glycerol oxygens at $X_{\text{DMSO}}=0$ mol\%.  
		(b) Survival probability $S(t)$ of water around the nitroxide radical oxygen. 
		(c) The mean escape time ($\langle\tau\rangle$) of water from the nitroxide radical oxygen of Tempo as a function of $X_{\text{DMSO}}$. 
		$\langle\tau\rangle$ is used to estimate the diffusion constant of water around Tempo. 
		\label{FigS_trap_water}}
\end{figure}

 \begin{figure}
\centering
 	\includegraphics[width=0.9\columnwidth]{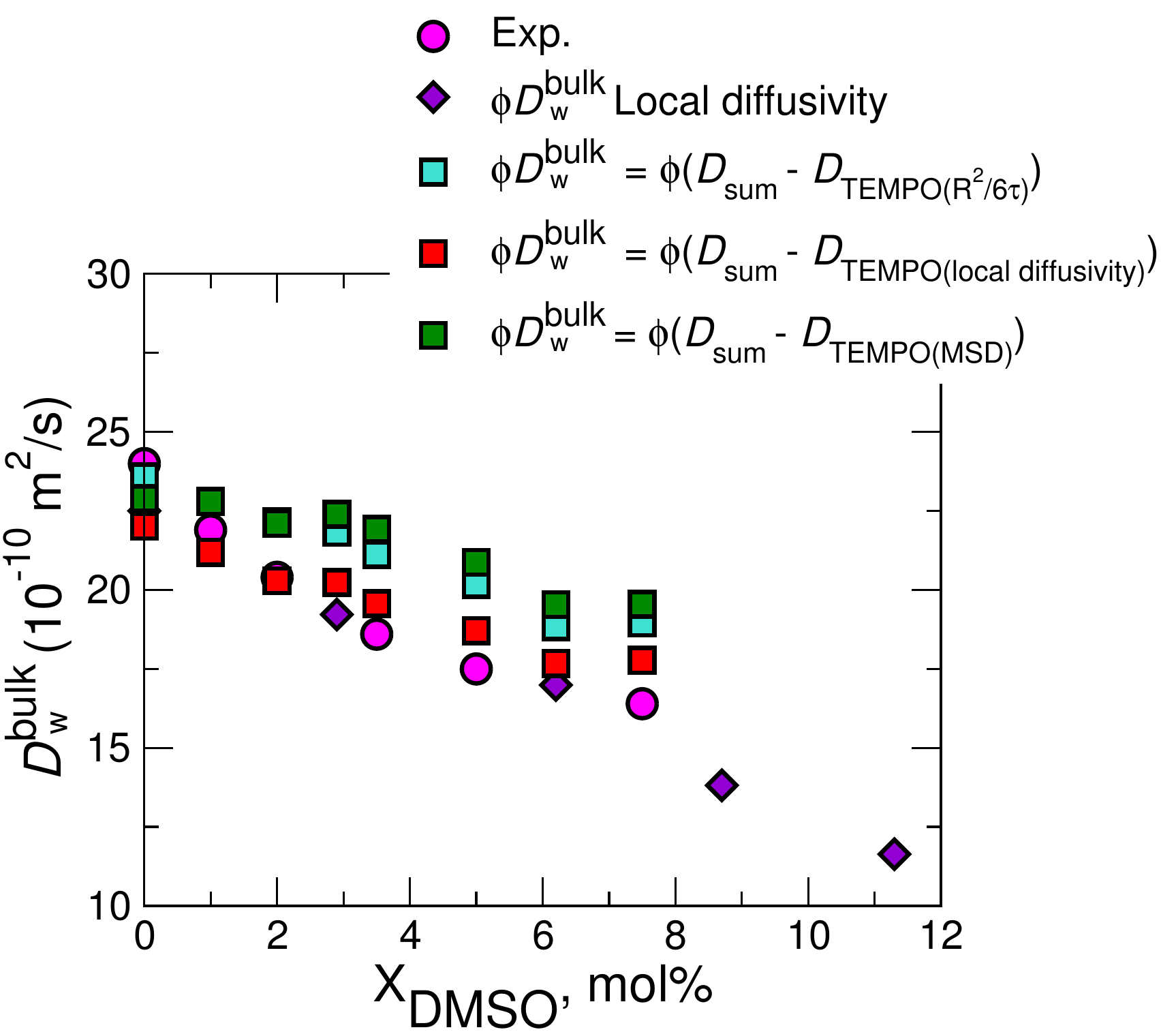}
  	\caption{
  		$D^{\text{bulk}}_{\text{w}}$ values using spin-label measurement \cite{cheng2015BJ},  
		calculated for pure water, around Tempo moiety in solution. 
		To calculate $D^{T,\text{bulk}}_{\text{w}}$ values around Tempo moiety from simulations,  
		we subtracted the contribution of Tempo from the total diffusion constant.   
  		 All the four different ways of calculating diffusion constant of bulk water 
  		 (using Eq.3, Eq.2, mean square displacement) give results comparable to each other.
		 \label{Comparison_DTempo_three_methods}}
\end{figure}

\begin{figure*}[ht]
\centering
\includegraphics[width=2.0\columnwidth]{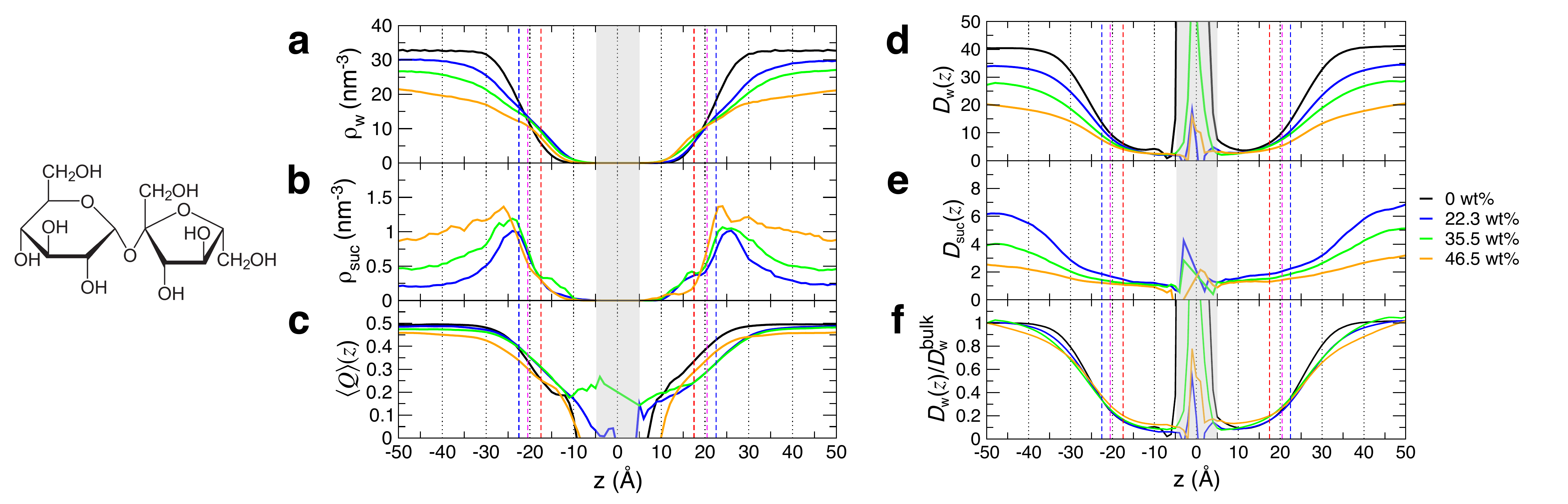}
\caption{Effects of sucrose on water molecules on bilayer surfaces at various sucrose weight percent $X_{\text{sucrose}}=0$, 22.3, 35.5, 46.5 wt\%. 
Density profiles of {\bf (a)} water and {\bf (b)} sucrose. 
{\bf (c)} Tetrahedral order parameter. 
Local diffusivities ($\times 10^{-10}$ m$^2$/s) of {\bf (d)} water and {\bf (e)} sucrose.  
{\bf (f)} Water diffusivity normalized by the bulk diffusion constant.   
\label{sucrose}}
\end{figure*}

\begin{figure*}
\centering
\includegraphics[width=1.7\columnwidth]{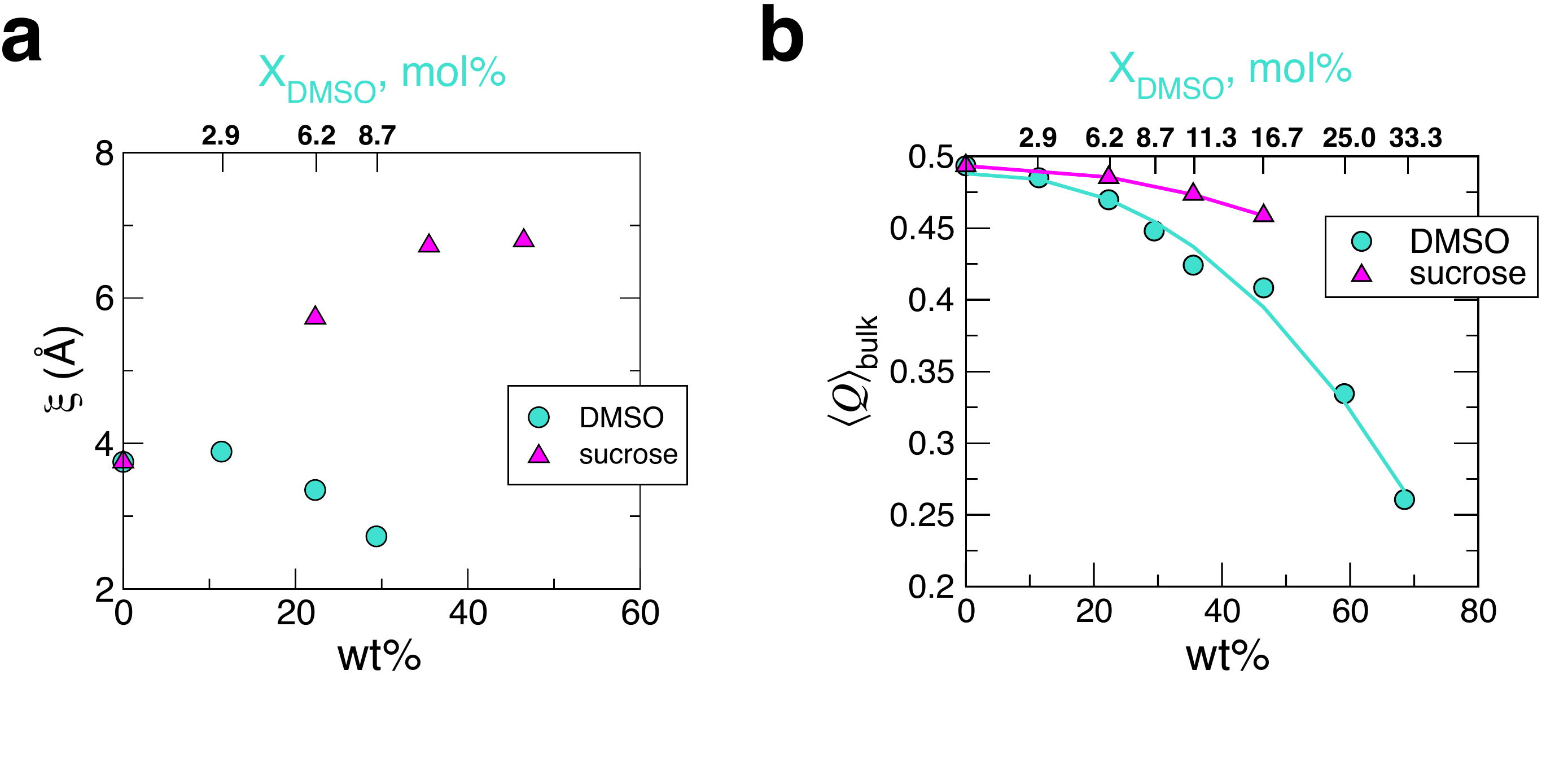}
\caption{Disparate effects of DMSO and sucrose on
water structure and dynamics. 
(a) The width ($2\xi$) of
solvent-bilayer interface as a function of cosolvent concentration (weight \%). Each $\xi$ was obtained by fitting the density profile to Eq.\ref{eqn:Cahn-Hillard}.
(b) Comparison of tetrahedral order parameters probing the water structure at varying weight \% of cosolvents (DMSO and sucrose) in the bulk. 
DMSO is more efficient than sucrose in disrupting the tetrahedral geometry of water H-bond network. 
\label{xi_Tetrahedral}}
\end{figure*}

\begin{figure}
\centering
	\includegraphics[width=0.9\columnwidth]{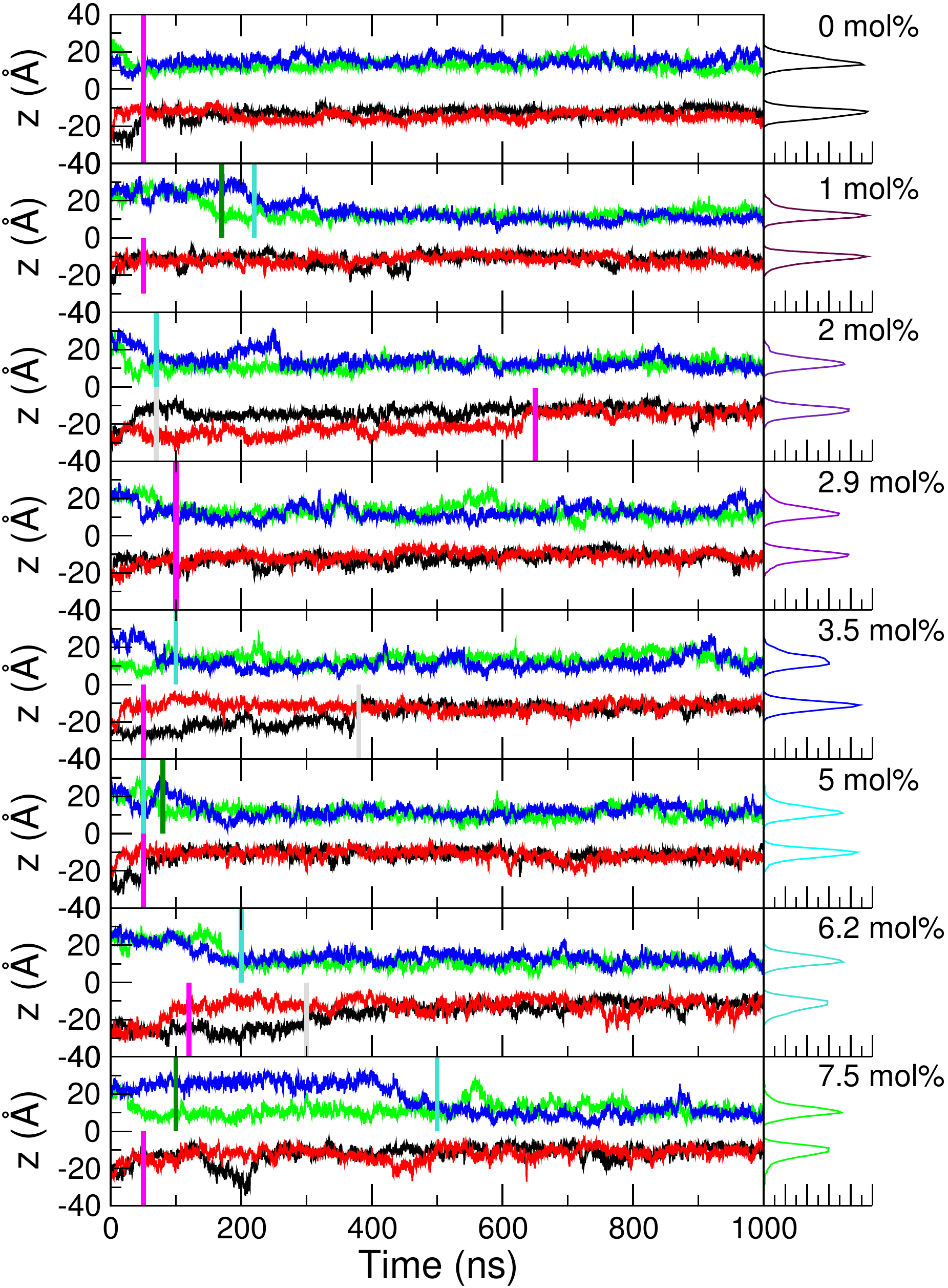}
	\caption{
		Dynamics of Tempo moiety attached to the head group at various DMSO concentrations are probed using the position of the nitroxide oxygen along the $z$-axis. 
		Time trajectories obtained from the four Tempo moieties are shown in different colors (blue, green from the upper leaflet, and black, red from the lower leaflet). 
		Note the position of Tempo moiety undergoes large fluctuation over time. 
		Surface water diffusion constants were evaluated using the simulation data collected after the Tempo-PC probe was relaxed from the initial position, reaching the steady state dynamics, which is specified with the bars in each graph. 
		The lifetime analysis of water around Tempo in Fig.\ref{FigS_trap_water} was conducted for the time intervals after these bars. 
		\label{Tempo_pc_z_trace_with_histo}}
\end{figure}

\begin{figure*}
\centering
	\includegraphics[width=2.0\columnwidth]{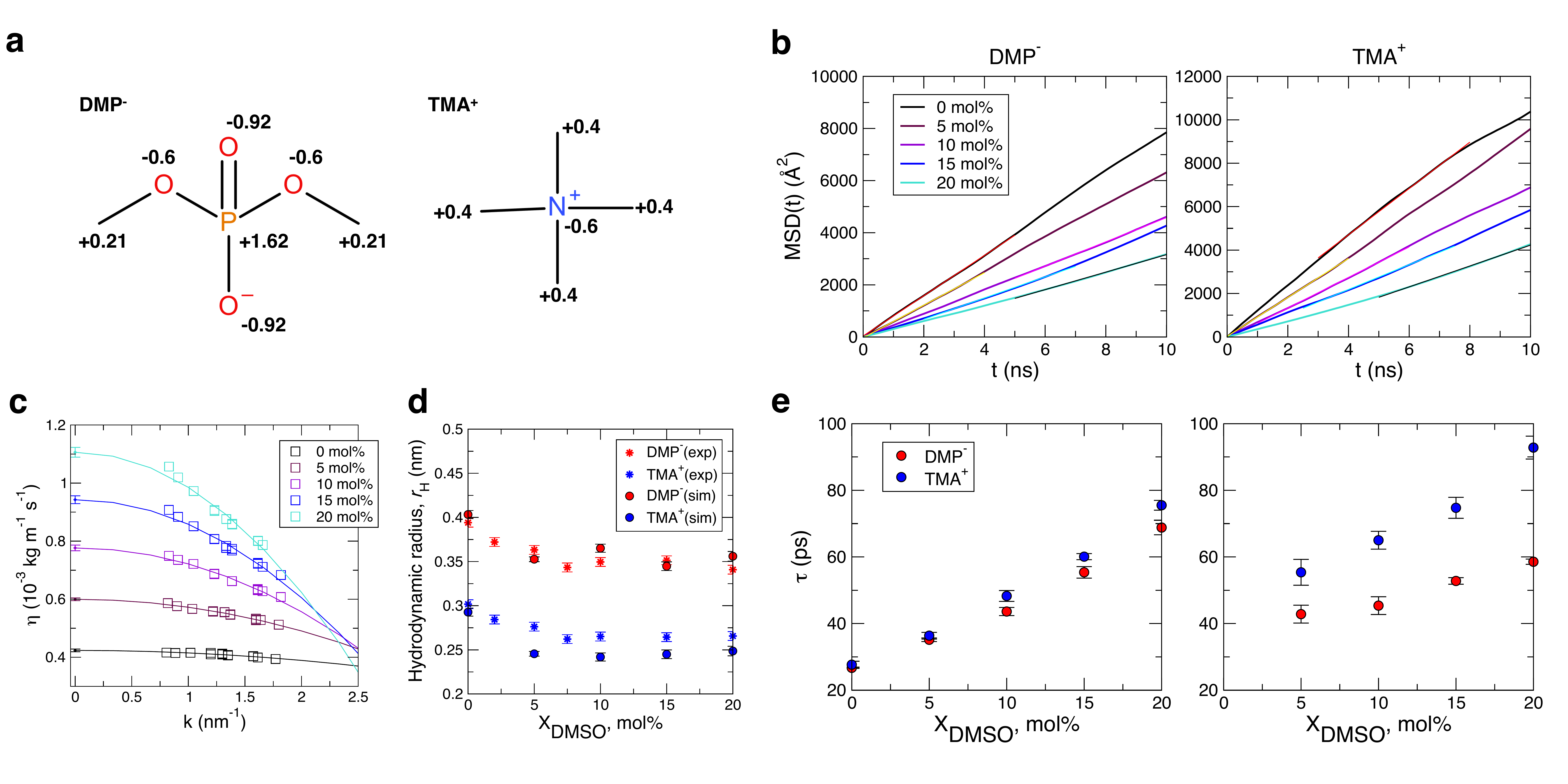}
	\caption{
		Simulations of PC head groups, dimethyl phosphate (DMP$^-$) and tetramethylammonium (TMA$^+$), at various DMSO concentrations (0 -- 20 mol\%) to estimate their hydrodynamic radii $r_{\rm{H}}$ in the bulk. 
		Simulation results using Berger lipid force field essentially reproduce the experimental results of $r_{\rm{H}}$, reported by Schrader \emph{et al.}\cite{Schrader2015PNAS}.
		(a) Chemical structure and partial charges of DMP$^-$ and TMA$^+$.
		The force field parameters of DMP$^-$ and TMA$^+$ except for the partial charges are based on the parameters of Berger lipid force field.  
		The partial charges of DMP$^-$ were taken from OPLS force field \cite{price2001JCC} that has the closest charge composition with Berger lipid force field.
		For TMA$^+$, the charge of nitrogen atom was modified from $-0.5$ to $-0.6$ in accord with the standard CHARMM force fields for TMA$^+$ \cite{mackerell98JPCB}, so as to adjust the net charge to be $+1$.   
		(b) Time-averaged mean square displacement (MSD) of phosphorus atom in DMP$^-$ and nitrogen atom in TMA$^+$ to calculate diffusion constant of PC head group.
		The slopes of MSD, depicted as the lines with different colors, were obtained by linear fits.
		(c) Solvent viscosities of DMSO-H$_2$O systems were obtained using transverse-current autocorrelation-function (TCAF) calculation \cite{palmer1994PRE}.
		Additional DMSO-H$_2$O mixture systems at various DMSO concentrations were also simulated to predict the solvent viscosity. 
		To obtain the viscosity at infinite wavelength, the $k$-dependent viscosities are fitted with $\eta(k) = \eta_0 (1 - a k^2)$ where $\eta_0$ is the infinite system limit of $\eta$.
		The resulting fit and estimated viscosity ($\eta_0$) at $k=0$ is given by solid line and diamond symbol with error bar, respectively.  
		(d) Hydrodynamic radii ($r_H$) of DMP$^-$ and TMA$^+$ calculated in the bulk solution with increasing $X_{\text{DMSO}}$. 
		The calculated diffusion ($D$) and viscosity ($\eta$) values were used to compute $r_{\rm{H}}$ of the solutes in the DMSO-H$_2$O mixtures, based on the Stokes-Einstein relation.
		$D=\frac{k_{\text{B}}T}{6 \pi \eta r_{\rm{H}}}$, where $k_{\text{B}}$ is Boltzmann's constant and $T$ is the temperature. 
		As a result, the calculated viscosity values are lower and diffusion constants are faster than experimentally obtained values.
		When the two values are multiplied to yield $r_H=k_BT/6\pi\eta D$, The $r_{\rm{H}}$ values, obtained by multiplying the two aforementioned values, are in excellent agreement with the radii acquired from PFG NMR measurements ($\ast$) \cite{Schrader2015PNAS}. 
		(e) $X_{\text{DMSO}}$-dependent lifetimes of water (left) and DMSO (right) in the first solvation shell around DMP$^-$ and TMA$^+$. 
		\label{FigS_DMP_TMA}}
\end{figure*}

\end{document}